\documentclass[fleqn,usenatbib]{mnras}

\usepackage{mathptmx}

\usepackage[T1]{fontenc}
\usepackage{soul}

\DeclareRobustCommand{\VAN}[3]{#2}
\let\VANthebibliography\thebibliography
\def\thebibliography{\DeclareRobustCommand{\VAN}[3]{##3}\VANthebibliography}

\usepackage{graphicx}	
\usepackage{amsmath}	
\usepackage{amssymb}	
\usepackage{mathtools}
\usepackage{float}
\usepackage{graphicx}
\usepackage{ulem}
\usepackage{lscape}
\usepackage{xspace}

\title[A road-map to white dwarf pollution]{A road-map to white dwarf pollution: Tidal disruption, eccentric grind-down, and dust accretion}

\author[Brouwers Et Al.]{
	Marc G. Brouwers,$^{1}$\thanks{E-mail: mgb52@cam.ac.uk}
	Amy Bonsor$^{1}$ and Uri Malamud$^{2,3}$
	\\
	$^{1}$Institute of Astronomy, University of Cambridge, Madingley Road, Cambridge CB3 0HA \\
	$^{2}$Department of Physics, Technion - Israel Institute of Technology, Technion City, 3200003 Haifa, Israel\\
	$^{3}$School of the Environment and Earth Sciences, Tel Aviv University, Ramat Aviv, 6997801 Tel Aviv, Israel
}

\date{Accepted XXX. Received YYY; in original form ZZZ}

\pubyear{2021}

\begin{document}
	\maketitle
	
	\begin{abstract}
	    A significant fraction of white dwarfs show metal lines indicative of pollution with planetary material but the accretion process remains poorly understood. The main aim of this paper is to produce a road-map illustrating several potential routes for white dwarf pollution and to link these paths to observational outcomes. Our proposed main road begins with the tidal disruption of a scattered asteroid and the formation of a highly eccentric tidal disc with a wide range of fragment sizes. Accretion of these fragments by Poynting-Robertson (PR) drag alone is too slow to explain the observed rates. Instead, in the second stage, several processes including differential apsidal precession cause high-velocity collisions between the eccentric fragments. Large asteroids produce more fragments when they disrupt, causing rapid grind-down and generating short and intense bursts of dust production, whereas smaller asteroids grind down over longer periods of time. In the final stage, the collisionally produced dust circularises and accretes onto the white dwarf via drag forces. We show that optically thin dust accretion by PR drag produces large infrared (IR) excesses when the accretion rate exceeds $10^7$ g/s. We hypothesise that around white dwarfs accreting at a high rate, but with no detected infrared excess, dust circularisation requires enhanced drag - for instance due to the presence of gas near the disc’s pericentre.
	\end{abstract}
	
	\begin{keywords}
		white dwarfs – planetary systems - transients: tidal disruption events - planet–disc interactions
	\end{keywords}
	

\section{Introduction}\label{sect:introduction}
Over the last decade, it has become clear that planetary systems around solar mass stars are the norm rather than the exception, with estimated occurrence rates around unity \citep{Berta2015, Dressing2015, Mulders2019, Zhu2021}. When these stars exit the main-sequence, subsequent stellar mass-loss leads to an expansion of the surrounding orbits, protecting material outside a few AU from the increased stellar flux \citep{Veras2011, Veras2012, Veras2016a}, while any objects within are engulfed \citep{Mustill2012, Villaver2014}. The enduring presence of planetary material around white dwarfs is inferred from the fraction of 25\% - 56\% of systems that show metal lines in their spectra \citep{Zuckerman2003, Zuckerman2010, Koester2014, Wilson2019}. So far, the presence of 21 different heavy elements has been detected in white dwarf photospheres, with 19 of these found in the single system GD 362 \citep{Zuckerman2007, Xu2013, Xu2017, Melis2017}.
\begin{figure*}
\centering
\includegraphics[width=\hsize]{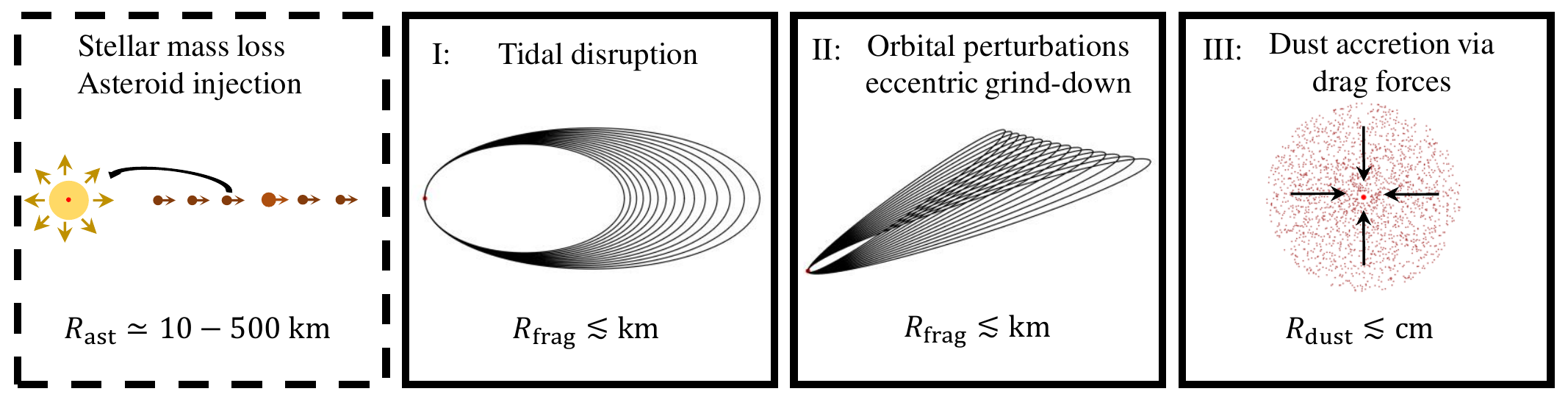} 
\caption{Schematic diagram of the main road to white dwarf pollution examined in this work. In the precursor to its pollution, the star sheds its outer layers during post-main sequence evolution (dashed border). This widens and destabilises the orbits of surrounding bodies and causes some asteroids to be scattered towards the star through interactions with nearby planets. In the first stage of accretion, asteroids that cross the Roche radius are tidally disrupted and form eccentric structures with an orbital spread, which we refer to as (eccentric) \textit{tidal discs}. The orbits of the surviving fragments are perturbed via various processes, including differential precession, causing high-velocity collisions and grind-down within the eccentric tidal disc (stage II). The resulting dust then circularises and accretes onto the star via various drag forces (stage III). \label{fig:sketch_phases}}
\end{figure*}

Detailed analysis of these polluted white dwarfs has opened a new channel for the investigation of exoplanetary systems. In particular, the detection of multiple heavy elements provides an opportunity to directly study the composition of planetary material outside the Solar System. In most cases, both the abundances \citep{Xu2014b, Xu2019, Wilson2015} and oxygen fugacity \citep{Doyle2020,Doyle2021} are found to match bulk Earth to zeroth order, although core-rich, mantle-rich and even crust-dominated objects have also been observed \citep{Hollands2017, Swan2019b, Hollands2021}. These systems have been interpreted as evidence for early planetsimal formation followed by collisional processing \citep{Harrison2018, Harrison2021} and indicate that the majority of exo-planetesimals are differentiated \citep{Bonsor2020}. In future models, accounting for pressure-sensitivity of different elements will allow the origin of the accreted objects to be determined as well (Buchan et al. in prep).

However, while the study of polluted white dwarfs has blossomed into a field of its own, the processes via which planetary material is accreted remain poorly understood. We know that the initial trigger for pollution is likely to be stellar mass loss, which widens planetary orbits and strengthens the interactions between planets. This can destabilise tightly-packed planetary systems, even if they were previously stable \citep{Debes2002, Maldonado2020, Maldonado2021}, while systems with more space between the planets likely survive intact \citep{Duncan1998, Veras2016b}. If the planetary system contains an asteroid belt, a single massive planet orbiting interior to the belt can scatter large numbers of asteroids within its expanding chaotic zone \citep{Bonsor2011, Mustill2018}. An outer planet can be similarly effective and scatter asteroids around expanding interior mean motion resonances \citep{Debes2012a, Antoniadou2016}. Asteroids or planets that pass within the Roche radius tidally disrupt into highly eccentric discs whose shapes range from narrow if the asteroid was small \citep{Veras2014,Veras2021a, Nixon2020}, to wider and even partially unbound if the object was terrestrial-sized \citep{Malamud2020a, Malamud2020b}. 

From this point on, the evolution of the fragments remains more obscured. As an end-point of their evolution, observations show that a minority of polluted white dwarfs are surrounded by dust \citep{Rocchetto2015, Farihi2016, Wilson2019}, whose emission typically varies within several years at mid-infrared \citep{Farihi2018, Swan2019a, Swan2020}, but less commonly at near-infrared \citep{Rogers2020}. Some of these systems also show evidence of on-going gas production \citep{Gaensicke2006, Gaensicke2007, Gaensicke2008, Manser2020} and they occasionally contain larger, transiting bodies \citep{Vanderburg2015, Manser2019b, Vanderbosch2020, Vanderbosch2021, Guidry2021}. While no complete description of the accretion process currently exists, the general hypothesis is that small fragments are circularised by Poynting-Robertson (PR) drag \citep{Rafikov2011a, Veras2015a, Veras2015b}, while larger fragments require a prior phase of collisional grind-down \citep{Jura2003, Jura2007, Wyatt2011, Li2021}. Other mechanisms that induce orbital changes after disruption are the Yarkovski force \citep{Veras2015a, Veras2015b, Malamud2020b}, potential interactions with pre-existing material around the star \citep{Grishin2019,Malamud2021} and magnetic Alfv\'en-wave drag \citep{Zhang2021}.

In this paper, we systematically investigate how planetary material accretes onto white dwarfs. We eventually produce a road-map illustrating several potential routes for white dwarf pollution with links to observational outcomes (Fig. \ref{fig:overview}). The main path to accretion that we examined (see Fig. \ref{fig:sketch_phases}) begins with the tidal disruption of a scattered asteroid to form a highly eccentric tidal disc (Sect. \ref{sect:disruption}). We evaluate its morphology and constrain the upper and lower bounds of their fragment sizes due to radiative and tidal forces. Then, we present a short intermezzo where we consider the merits and limitations of a simple collision-less evolution model via PR drag (Sect. \ref{sect:collisionless}), which we find cannot drive sufficiently rapid accretion, even under the most optimistic assumptions. We continue our main road to pollution in Sect. \ref{sect:perturbations}, where we discuss how various processes induce high-velocity collisions between fragments. These collisions take place while the fragments still travel along the highly eccentric orbits ($e>0.999$) on which they are released. This notion is fundamentally different from previous models that calculated collisions between fragments that were already supposed to have circularised \citep{Kenyon2017a, Kenyon2017b, Swan2021}. In Sect. \ref{sect:collisional_model}, we present a simple but quantitative calculation of eccentric collisional grind-down based on the fragment's differential rates of apsidal precession, a process that likely induces collisions in all tidal discs on this scale. We then enter the third stage of our accretion scenario where we model the geometry and emission of the dust that is produced (Sect. \ref{sect:IR_Excess}). We discuss variations of this model as well as alternative paths to white dwarf pollution in Sect. \ref{sect:discussion}, including observational outcomes when they are sufficiently well understood. Finally, we conclude our work in Sect. \ref{sect:conclusions}

\section{Stage I: from asteroid to tidal debris disc}\label{sect:disruption}
\subsection{The tidal disruption criterion}
\begin{figure}
\centering
\includegraphics[width=\hsize]{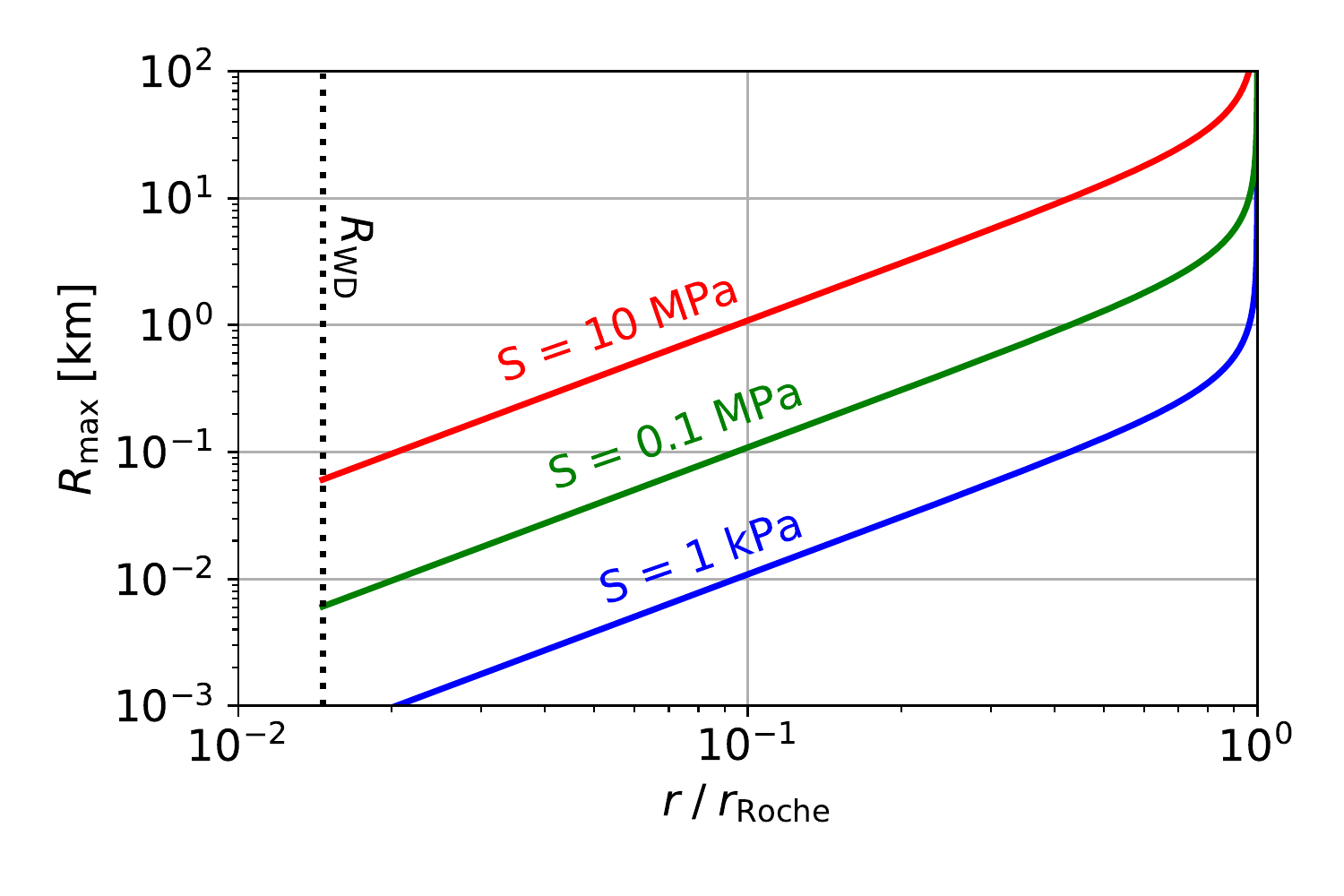} 
\caption{The maximum sizes of objects that are safe from tidal disruption at a distance $r$ from a 0.6 $\mathrm{M_\odot}$ white dwarf, plotted for three different tensile strengths. Large asteroids (>100 km) always disrupt close to the Roche radius of the star, whereas smaller fragments ($\ll$ km) can survive deep pericentre passages ($r\ll r_\mathrm{Roche}$) due to their material strength. \label{fig:Strengths}} 
\end{figure}

The outer edge of the disruption zone is set by the distance where an asteroid's internal strength and self-gravity are overcome by tidal forces. The details of this process depend on the asteroid's shape and composition, as well as on its path and potential rotation \citep{Dobrovolskis1982, Dobrovolskis1990, Davidsson1999, Davidsson2001}. We content ourselves here by by considering an idealised case of breakup by tensile failure, likely the most common type of tidal disruption for solid bodies. We adopt a similar approach as \citet{Bear2015} and \citet{Brown2017} and identify the breakup criterion for a spherical, non-rotating asteroid of size $R_\mathrm{ast}$ and density $\rho_\mathrm{ast}$ as the point where the summed forces from the tensile strength ($F_\mathrm{S}$) and self-gravity ($F_\mathrm{SG}$) first fail to compensate the tidal force induced by the gravitational gradient ($F_\mathrm{T}$):
\begin{equation}\label{eq:force_balance}
    F_\mathrm{S} + F_\mathrm{SG} + F_\mathrm{T} \simeq -S R_\mathrm{ast}^2 - \frac{GM_\mathrm{ast}^2}{R_\mathrm{ast}^2} + \frac{2GM_\mathrm{WD}M_\mathrm{ast}R_\mathrm{ast}}{r^3}= 0,
\end{equation}
where $G$ is the gravitational constant, is $r$ is the distance to the white dwarf and S is the material's tensile strength. The masses of the white dwarf and the asteroid are indicated by $M_\mathrm{WD}$ and $M_\mathrm{ast}$, respectively. In the gravity-dominated regime ($|F_\mathrm{SG}| \gg |F_\mathrm{S}|$), the distance at breakup is mostly independent of the asteroid's size and occurs at the classical Roche radius \citep[e.g.]{Davidsson1999,Bear2013, Veras2014, Malamud2020a,Malamud2020b}:
\begin{equation}\label{eq:r_roche_ast}
    r_\mathrm{Roche} =  {\left(\frac{ 2 \rho_\mathrm{WD}}{\rho_\mathrm{ast}}\right)}^\frac{1}{3} R_\mathrm{WD}.
\end{equation}
To quantify Eq. \ref{eq:r_roche_ast}, it is necessary to specify the white dwarf density $\rho_\mathrm{WD}$. Neglecting the slight temperature dependence, this relationship can be approximated by \citep{Nauenberg1972}:
\begin{equation}\label{eq:rho_wd}
    R_\mathrm{WD} = 0.0127 \; \mathrm{R_\odot} \left(\frac{M_\mathrm{WD}}{\mathrm{M_\odot}}\right)^{-1/3} \left(1-0.607 \left(\frac{M_\mathrm{WD}}{\mathrm{M_\odot}}\right)^{4/3}\right)^\frac{1}{2}.
\end{equation}
The Roche radius amounts to roughly $1 \; \mathrm{R_\odot}$ for a 0.6 $\mathrm{M_\odot}$ white dwarf. If they are monolithic (as opposed rubble-pile aggregates, see below), smaller asteroids can survive a certain distance within the Roche radius until the extra barrier of their internal strength is overcome. Accounting for this, Eq. \ref{eq:force_balance} can be solved to yield a maximum object size ($R_\mathrm{max}$) that can survive at a distance $r$ from a WD:
\begin{equation}\label{eq:R_max}
    R_\mathrm{max} = \frac{3}{4\pi \rho_\mathrm{ast}}\left(\frac{S}{G\left[\left(\frac{r_\mathrm{Roche}}{r}\right)^3-1\right]}\right)^\frac{1}{2}.
\end{equation}
Chondrite asteroid samples indicate that the upper end of tensile strengths is around 0.1-10 MPa \citep{Scheeres2015, Veras2020a, Pohl2020} but it is understood to vary by orders of magnitude depending on an asteroid's formation history and composition. For instance, estimates from modeling of cometary material composed of ice-coated interstellar silicate grains indicate sub-kP strengths \citep{Greenberg1995, Davidsson1999, Gundlach2016}. A formation via the gentle sticking of constituent particles can even result in so called rubble piles with near-zero effective strength. Such an object is for instance invoked to explain the rapid breakup of Shoemaker-Levy 9 in Jupiter's outer envelope \citep{Asphaug1994}. To visualise these differences, we plot the maximum intact object size as a function of distance for a range of tensile strengths in Fig. \ref{fig:Strengths}. The figure illustrates the dichotomy that arises based on the asteroid's size. Large (> 100 km) asteroids always fragment at the Roche radius regardless of their tensile strength. In contrast, because the tidal force scales as $F_\mathrm{T} \propto R^4$ and $F_\mathrm{S} \propto R^2$, smaller km-sized granite rocks can reach as close as a few percent of the Roche radius before they are torn apart.

\subsection{Tidal debris disc morphology}\label{sect:stream_morphology}
As the main body begins to break up, its fragments stray from the initial orbit and spread out over a range of energies. \citet{Malamud2020a, Malamud2020b} simulated this process for terrestrial-sized bodies and showed how the breakup typically proceeds in stages. Pericentre passages that are close to the Roche radius typically result in partial disruption and require several orbits in order to completely destroy the object. Each passage can add spin to the surviving object and progressively weaken it. The fragments that break off begin to form a stream that can gravitationally collapse perpendicular to its direction of motion. After several orbits, this produces a fully formed tidal disc of interlaced elliptical annuli.

While the details of the breakup process lead to some variation in the tidal disc, its basic morphological features agree with the simple impulse approximation method. In this framework, the disruption is assumed to occur instantaneously at a location $r_\mathrm{B}$ from the star, which for simplicity we take at the pericentre of the object's orbit. As the object disintegrates, its fragments are no longer guided by the centre of mass and continue on orbits corresponding to their energy at the point of separation. The ones facing the white dwarf are more gravitationally bound than their velocity warrants and move to a tighter orbit, while those on the other side of the asteroid migrate away from the white dwarf. Neglecting the small effect of binding energy in asteroid-sized objects, their specific energy ($\epsilon_\mathrm{i}$) can be expressed as a sum of the kinetic ($\epsilon_\mathrm{k,i}$) and potential ($u_\mathrm{G,i}$) parts:
\begin{subequations}
\begin{align}
    \epsilon_\mathrm{i} &= \epsilon_\mathrm{k,i} + u_\mathrm{G,i}\\
    &= \frac{1}{2} GM_\mathrm{WD}\left(\frac{2}{r_\mathrm{B}}-\frac{1}{a_0}\right) - \frac{GM_\mathrm{WD}}{r_\mathrm{i}}, \label{eq:e_width}
\end{align}
\end{subequations}
where $r_\mathrm{B}$ is the breakup distance of the asteroid's centre to the white dwarf and $r_\mathrm{i} = r_\mathrm{B} + x R_\mathrm{ast}$ is the distance corresponding the individual fragments (with $-1<x<1$) and $a_0$ is the asteroid's semi-major axis. Correspondingly, the fragment's new semi-major axes ($a_\mathrm{i}$) are spread along
\begin{subequations}
\begin{align}
    a_\mathrm{i} &= -\frac{GM_\mathrm{WD}}{2\epsilon_\mathrm{i}} \label{eq:a_new_1} \\
    &= a_0 {\left(1+ 2a_0\frac{r_\mathrm{B}- r_\mathrm{i}}{r_\mathrm{B}r_\mathrm{i}}\right)}^{-1}, \label{eq:a_new_2}
\end{align}
\end{subequations}
with eccentricities ($e_\mathrm{i}$) equal to
\begin{equation}\label{eq:e_orbits}
    e_\mathrm{i} = \left(1-\frac{r_\mathrm{i}}{a_\mathrm{i}}\right).
\end{equation}
The asteroid moves parallel to the plane of motion of its centre-of-mass prior to its breakup. This means that fragments get imparted an inclination ($i_\mathrm{i}$) depending on their vertical position that varies between $0<i<R_\mathrm{ast}/r_\mathrm{i}$. With this, the tidal debris discs have an approximate height of $2R_\mathrm{ast}$ at pericentre, which grows to many times the size of the body at apocentre, where it can be estimated at $H_\mathrm{apo} \sim 4 R_\mathrm{ast} a_0 / r_\mathrm{B}$. In our subsequent modeling, we take the inclination constant as a function of time, although the vertical evolution of these tidal discs currently remains poorly understood.
\begin{figure}
\centering
\includegraphics[width=\hsize]{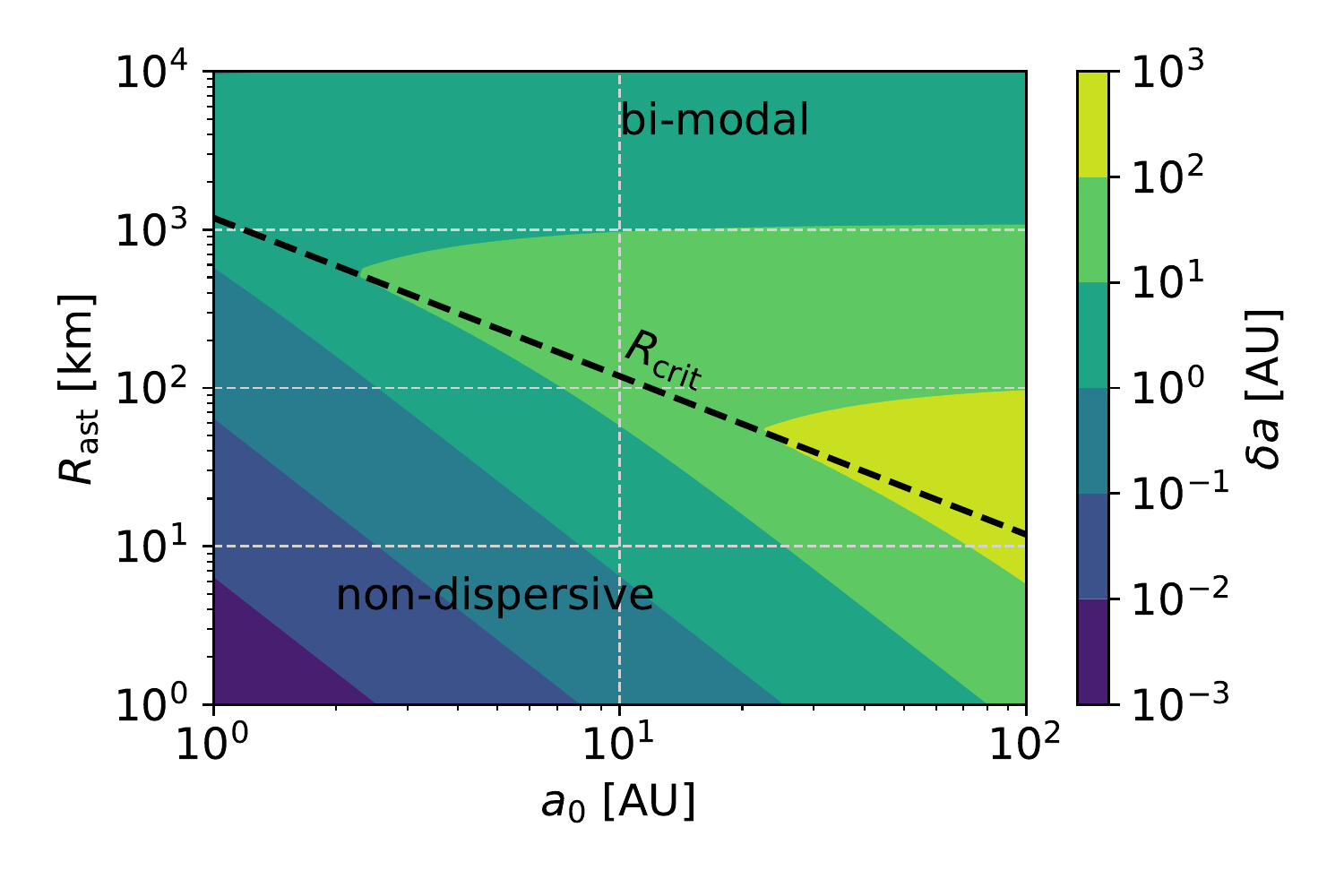} 
\caption{Orbital width ($\delta a$) that contains 90 \% of the bound fragments around a 0.6 $\mathrm{M_\odot}$ white dwarf. Small asteroids on tight orbits experience non-dispersive breakup and form a non-dispersive stream. The width of the tidal disc increases with the progenitor size and its orbital separation. Very large objects like planets with $R \gg R_\mathrm{crit}$ experience bimodal breakup with half of their mass being ejected from the system. \label{fig:disruption_width}} 
\end{figure}

\begin{figure}
\centering
\includegraphics[width=\hsize]{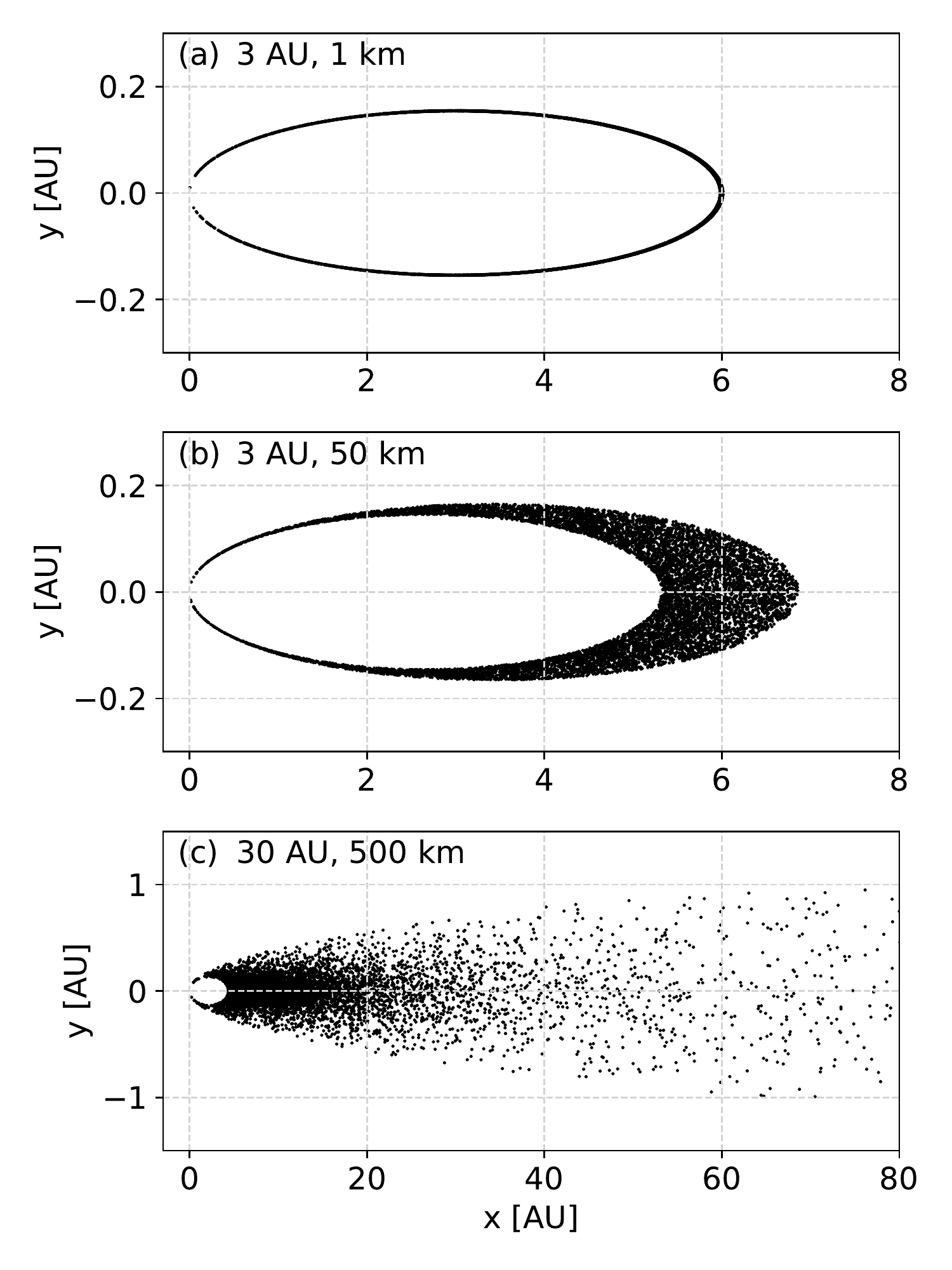} 
\caption{Post-breakup orbits after a strengthless asteroid is tidally disrupted near a 0.6 $M_\mathrm{\odot}$ white dwarf. Panel (a) indicates the tidal disc that forms when a small (1 km from 3 AU) asteroid disrupts, yielding a thin orbital spread. Panel (b) corresponds to a 50 km asteroid from 3 AU, which produces a wider - but still completely bound disc. When both the asteroid size and semi-major axis are increased to 500 km and 30 AU (c), the outcome is a bimodal disruption event, with nearly half of the material becoming unbound from the white dwarf. \label{fig:new_orbits}} 
\end{figure}

Depending on the size and semi-major axis of the asteroid prior to breakup, the tidal discs described as by Eqs. \ref{eq:a_new_1} - \ref{eq:e_orbits} form with various shapes. Asteroids that originate from wide orbits are only loosely bound to the star and when their size exceeds the threshold of $R_\mathrm{crit} \simeq \frac{r_\mathrm{B}^2}{2a_\mathrm{0}-r_\mathrm{B}}$ \citep{Malamud2020a}, some of its fragments become unbound after breakup ($a_\mathrm{i}<0$), with up to half of the material being expelled from the system in the most extreme case. These highly dispersive tidal disruption events have previously been linked to planetesimal seeding of other systems \citep{Rafikov2018}. The fragments that remain on bound orbits spread out over a range in semi-major axes, approximately distributed evenly in the energy range of Eq. \ref{eq:e_width} \citep{Malamud2020a}. Since the orbital energy scales as $a^{-1}$, most bound fragments become clustered on tight orbits. Therefore, we can use the inner ($a_\mathrm{in}$) and outer ($a_\mathrm{out}$) bound fragments to define an effective width $\delta a$ of the tidal disc as
\begin{equation}
    \delta a = \frac{\chi a_\mathrm{in}(a_\mathrm{out}-a_\mathrm{in})}{\chi a_\mathrm{in} + (1-\chi)a_\mathrm{out}},
\end{equation}
where $\chi$ is the mass fraction of bound fragments included in the width of the disc. We plot this effective disc width for a range of semi-major axes and asteroid sizes in in Fig. \ref{fig:disruption_width}, which shows the clear dichotomy between the non-dispersive and bimodal regimes. We also visualise three examples of the different regimes in Fig. \ref{fig:new_orbits}. The top panel (a) indicates the tidal disc that forms when a small asteroid (1 km from 3 AU) disrupts, similar to the models by \citet{Debes2012a}, \citet{Veras2014}, and \citet{Nixon2020}. As described by these authors, the result is a completely bound but spaghettified orbital structure. If the size of the asteroid and its semi-major axis are increased (panel b), the orbital band broadens until $\delta a >> a_0$ and almost half of the original material is ejected (panel c). In this bimodal regime, any further increase of the impactor size or semi-major axis begins to increase the concentration of orbits closer to the star and effectively reduces the width of the tidal disc (see top part of Fig. \ref{fig:disruption_width}).

\subsection{Debris size distribution}\label{sect:size_dist}
Immediately after the main body breaks up, its fragments become exposed to the same tidal force that broke up their progenitor. These fragments are smaller than the original body and thus require a lower tensile strength to resist further breakup and initially remain stable. However, if the asteroid was scattered onto an orbit with its pericentre closer to the star than its initial original breakup distance, the fragments experience an increased peak tidal force and may disrupt again. In this simple picture, the maximum fragment size $R_\mathrm{max}$ follows from Eq. \ref{eq:R_max} with $r=r_\mathrm{peri}$ and can be identified from Fig. \ref{fig:Strengths}. Because asteroids are most likely to be scattered to orbits with pericentres near the edge of the Roche radius \citep{Veras2021a}, it may be expected that the largest fragment is typically similar to the size of the asteroid itself. It is not necessary that this happens in practice, however, as tighter orbits or rubble pile asteroids composed of smaller constituent particles may not lead to any large surviving fragments. In addition, the fragments can be spun up, making them easier to break up \citep{Malamud2020b}. In our model, we take $R_\mathrm{max}$ as a free parameter due to the large associated uncertainty, with a baseline value of 1 km, corresponding to the size where fragments with a 1 kPa strength survive near the Roche radius. We assume the same density for the fragments as we do for the asteroid.

\begin{figure}
\centering
\includegraphics[width=\hsize]{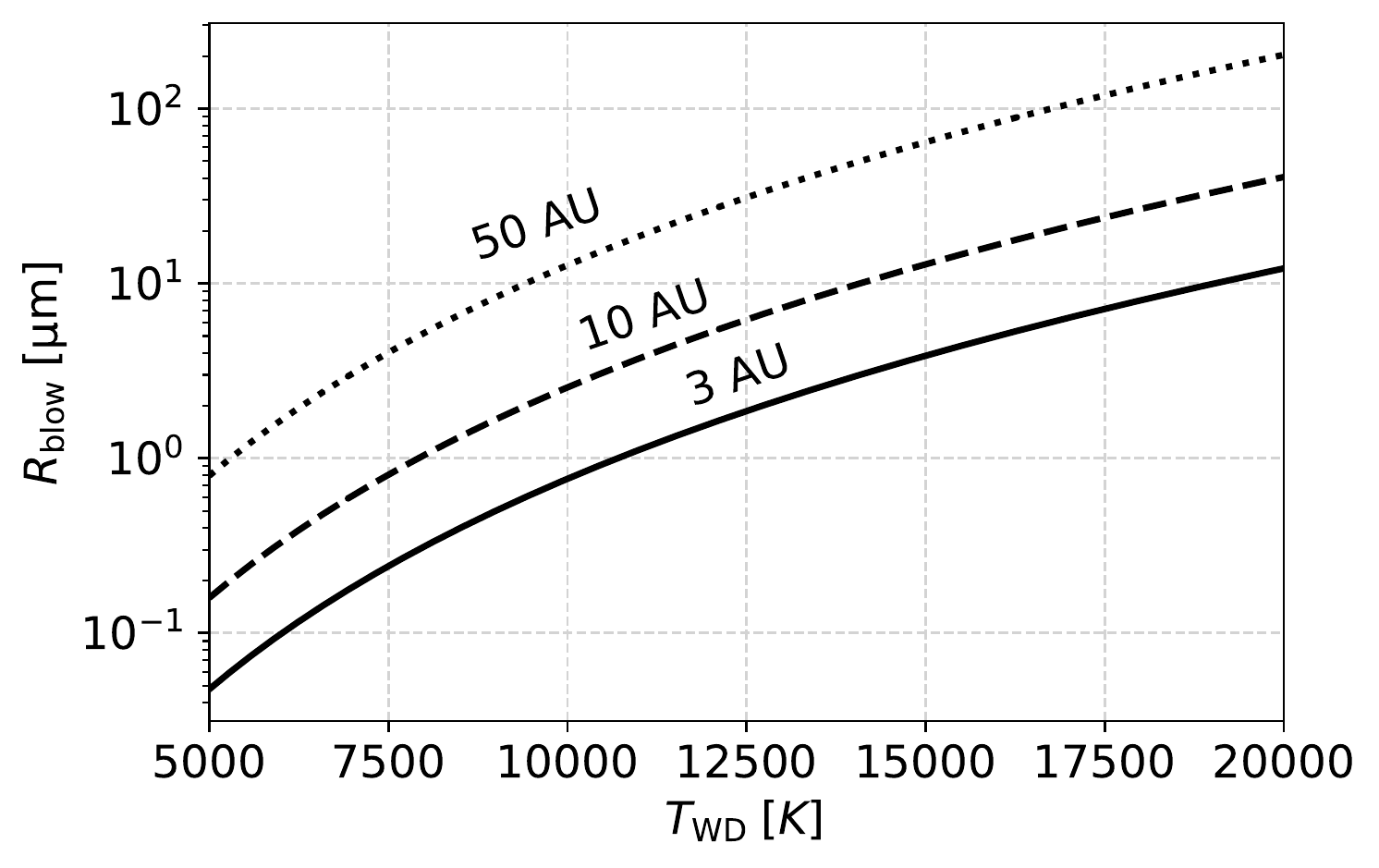} 
\caption{The smallest dust grains ($R_\mathrm{blow}$) that interact with the stellar light ($<Q>=1$) and can remain bound despite the star's radiation pressure. The figure assumes the disruption of a strengthless asteroid at the Roche radius of a 0.6 $\mathrm{M_\odot}$ white dwarf (from Eq. \ref{eq:R_blow}). The three lines correspond to different asteroid semi-major axes and the tidal disruption is assumed to occur at the orbit's pericentre. \label{fig:blowout}}
\end{figure}

On the other end of the distribution, the lower limit for bound fragments corresponds either to scale of the smallest dust grains that splinter off during breakup or to the blow-out size where radiation pressure from the stellar luminosity $L_\mathrm{WD}$ pushes the dust out of the system. This blow-out criterion provides an absolute lower limit to the fragment sizes and can be estimated from the ratio of radially oriented forces \citep{Burns1979}:
\begin{align}\label{eq:F_ratio}
    \frac{F_\mathrm{rad}}{F_\mathrm{G}} &= 0.0013 \; {\left(\frac{R}{\mathrm{\mu m}}\right)}^{-1}
    {\left(\frac{\rho_\mathrm{frag}}{2.7 \; \mathrm{g/cm^3}}\right)}^{-1}
    \left(\frac{L_\mathrm{WD}}{0.01\; L_\mathrm{\odot}}\right) \nonumber \\
    &
    {\left(\frac{M_\mathrm{WD}}{0.6 M_\mathrm{\odot}}\right)}^{-1}
    {\left(\frac{<Q>}{1}\right)},
\end{align}
where $<Q>$ is the radiation pressure coefficient, averaged over the stellar emission. While white dwarfs are not luminous enough to blow out micron-sized grains on circular orbits, the near-unity eccentricities of the fragments after the tidal disruption make these grains susceptible. We can estimate a typical blow-out size by taking the orbital parameters of the original asteroid with a breakup point at its pericentre. The smallest fragments are placed on unbound orbits when when $F_\mathrm{rad} /F_\mathrm{G} > 0.5(1-e)$, which, when combined with Eq. \ref{eq:e_orbits}, reduces to:
\begin{align}\label{eq:R_blow}
    R_\mathrm{blow} = 1.51 \;\mathrm{\mu m}\; \left(\frac{a_0}{\mathrm{AU}}\right) \left(\frac{r_\mathrm{B}}{R_\mathrm{\odot}}\right)^{-1} \left(\frac{L_\mathrm{WD}}{0.01 \; \mathrm{L_\odot}}\right) \nonumber \\
    \left(\frac{M_\mathrm{WD}}{0.6 \; \mathrm{M_\odot}}\right)^{-1} \left(\frac{\rho_\mathrm{frag}}{2.7 \; \mathrm{g/cm^3}}\right)^{-1},
\end{align}
assuming geometric scattering ($<Q> = 1$), which is valid for grains larger than the reduced peak stellar wavelength $\lambda_\mathrm{peak}/2\pi \simeq 0.02-0.1 \; \mathrm{\mu m}$, depending on the stellar temperature. We plot the blow-out size across a range of white dwarf temperatures for three asteroid semi-major axes in Fig. \ref{fig:blowout}. The figure indicates that the sizes typically range between 1-100 $\mathrm{\mu m}$ depending on the white dwarf temperature and the asteroid's orbit. We note, however, that if they are produced during the tidal disruption, a population of grains much smaller than $R_\mathrm{blow}$ can still remain remain in the system due to their reduced interaction with light at stellar wavelengths. We do not consider these tiny grains here because their properties also make them resistant to PR drag.

The distribution of fragments between $R_\mathrm{min}$ and $R_\mathrm{max}$ is determined by the fracture lines, the amount of sequential breakups and mergers, as well as the short collisional phase that follows \citep{Malamud2020a} and currently remains poorly constrained. We therefore opt to insert a truncated power-law for the fragment sizes:
\begin{equation}\label{eq:dm_dR_dist}
    \frac{dN}{dR} = C R^{-\alpha},
\end{equation}
where $\alpha$ is the scaling factor of the size distribution and the constant C is set by mass conservation. Because $R_\mathrm{max} \gg R_\mathrm{min}$ and assuming that $\alpha < 4$, it can also be written as
\begin{equation}\label{eq:mass_conservation}
    \frac{dM}{dR} = \frac{(4-\alpha)f_\mathrm{bound}M_\mathrm{ast}}{R_\mathrm{max}} {\left(\frac{R}{R_\mathrm{max}}\right)}^{3-\alpha},
\end{equation}
where $f_\mathrm{bound}$ is the mass fraction of post-breakup fragments that is bound to the white dwarf, a factor that follows from Eq. \ref{eq:a_new_2}. In the case of a scale-independent collisional cascade, the power law is characterised by $\alpha = 3.5$ and the mass is dominated by large fragments while the smaller particles dominate the cross section \citep{Dohnanyi1969, Tanaka1996, Wyatt2007, Wyatt2011}. We take this as our default value in our subsequently presented calculations but note that simulations of collisional cascades that account for scale-dependent effects suggest a slightly lower value of alpha. If the mass is instead more evenly distributed over the size bins, the value of $\alpha$ is closer to 3.
%
\section{Intermezzo: collision-less evolution via PR drag}\label{sect:collisionless}
Before we consider collisions between larger fragments, we first evaluate the potential of perhaps the simplest scenario, where fragments accrete via PR drag alone. We derive accretion rates as a function of the bounded size distribution and point out the limitations of this simple scenario.

\subsection{PR contraction timescale and accretion rate}\label{sect:pr_drag}
The main contribution from PR drag on a highly eccentric orbit occurs near the pericentre where the stellar flux is highest. In this calculation, we will make the a-priori assumption that the tidal disc is optically thin, which we later evaluate in Sect. \ref{sect:IR_Excess}. The averaged orbital equations of motion are described by \citet{Veras2015a, Veras2015b}:
\begin{subequations}
\begin{align}
    <\frac{da}{dt}>& = -\frac{3<Q>L_\mathrm{WD}(2+3e^2)}{16\pi\rho_\mathrm{frag}Rac^2(1-e^2)^\frac{3}{2}}, \label{eq:da_dt} \\
    <\frac{de}{dt}> &= -\frac{15<Q>L_\mathrm{WD}e}{32\pi\rho_\mathrm{frag}Ra^2c^2(1-e^2)^\frac{1}{2}}, \label{eq:de_dt}
\end{align}
\end{subequations}
where we again substitute $<Q>=1$ and use the simple relation
\begin{equation}\label{eq:L_WD} 
    L_\mathrm{WD} \simeq 3.26 \; L_\mathrm{\odot} \; {\left(0.1+\frac{t_\mathrm{WD}}{ \mathrm{Myr}}\right)}^{-1.18}
\end{equation}
\begin{figure}
\centering
\includegraphics[width=\hsize]{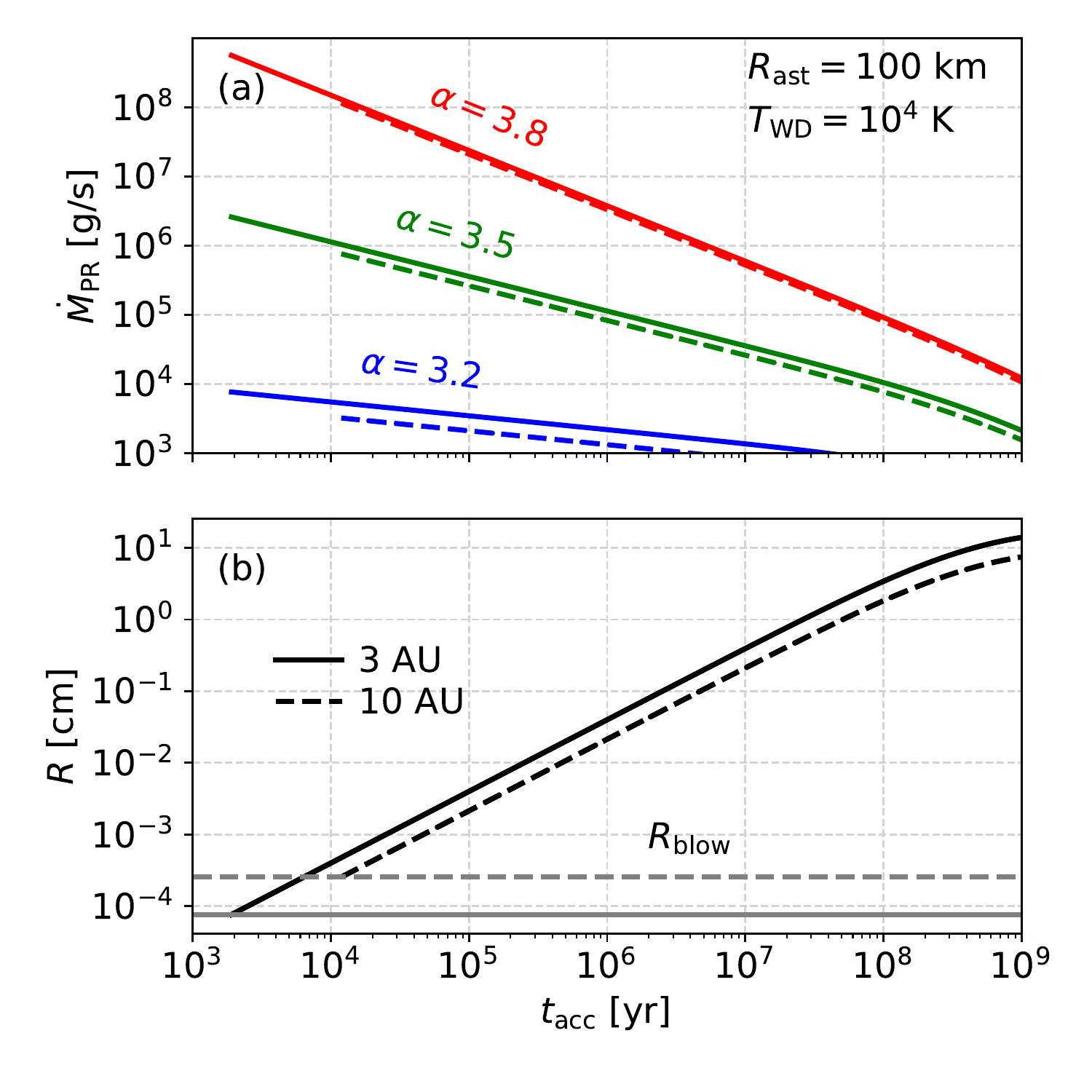} 
\caption{Collision-less mass accretion rates of asteroid fragments via PR drag onto a 0.6 $\mathrm{M_\odot}$ white dwarf. The top panel (a) shows how the accretion rate varies with the size distribution (different colors) and orbital separation (line style). The lower panel (b) indicates what fragments reach the star via PR drag after a certain time ($t_\mathrm{acc}$). Accretion starts with the smallest bound fragments $R_\mathrm{blow}$, as bigger fragments take longer to circularise. The curves flatten over time due to the declining luminosity of the cooling star (Eq. \ref{eq:L_WD}), meaning that fragments larger than $1-10$ cm cannot be accreted via PR drag alone. \label{fig:M_dot_analytical}}
\end{figure}
to relate the white dwarf's luminosity to its age $t_\mathrm{WD}$ based on Mestel theory \citep{Mestel1952}, with the same parameters as in \citet{Bonsor2010}. This prescription is valid up to 9 Gyr when the white dwarf undergoes crystallization and the cooling slows dramatically \citep{Althaus2010}. While more detailed cooling codes are available \citep{Salaris2013}, Mestel's relation captures the essential cooling trend for the first few Gyrs, which is sufficient for our purposes here.

The equations of motion (Eqs. \ref{eq:da_dt}-\ref{eq:L_WD}) are coupled and generally need to be solved numerically to obtain the accretion time as a function of fragment size $t_\mathrm{acc}(R)$. Because a white dwarf's luminosity remains approximately constant for a similar period of time as its age, a fragment's accretion time in the window $t_\mathrm{acc}<t_\mathrm{WD}$ is proportional to its size. For a size distribution $3<\alpha<4$, this leads to an accretion rate of:
\begin{subequations}
\begin{align}
    \dot{M}_\mathrm{PR} &= \frac{dM}{dR} {\left(\frac{dt_\mathrm{acc}}{dR}\right)}^{-1} \\
    &= \frac{(4-\alpha)f_\mathrm{bound} M_\mathrm{ast} t_\mathrm{acc}^{3-\alpha}}{ t_\mathrm{acc, max}^{4-\alpha}} \label{eq:analytical_mdot},
\end{align}
\end{subequations}
where $t_\mathrm{acc, max}$ is the accretion time of the largest fragment. We plot the PR accretion rates (accounting for stellar cooling) for 100 km asteroids that originate from 3 and 10 AU in Fig. \ref{fig:M_dot_analytical}, assuming three different values of $\alpha$. In agreement with Eq. \ref{eq:analytical_mdot}, the accretion rate declines as a function of time unless $\alpha<3$. The smallest fragments, therefore, typically determine the peak accretion rate in a collision-less scenario, even if most of the mass is contained in large fragments. The steeper the fragment size distribution, the higher the peak accretion rate and the steeper its decline as a function of time. Furthermore, we find that the accretion rate is only marginally dependent on the orbital parameters of the fragments. For a given value of $\alpha$, the PR accretion rate can also be be written as a simple scaling function. For the standard case of $\alpha =3.5$, this is:
\begin{equation}\label{eq:Mdot_PR_35}
\begin{split}
    \dot{M}_\mathrm{PR, 3.5} &\simeq 2.3 \cdot 10^{6} \; \mathrm{g/s} \; 
    \left(\frac{t_\mathrm{acc}}{10^4 \; \mathrm{yr}}\right)^{-\frac{1}{2}}
    \left(\frac{L_\mathrm{WD}}{0.01\; \mathrm{L_\odot}}\right)
    \left(\frac{R_\mathrm{max}}{1\; \mathrm{km}}\right)^\frac{1}{2} \\
    & \left(\frac{R_\mathrm{ast}}{100\; \mathrm{km}}\right)^3
    \left(\frac{f_\mathrm{bound}}{1}\right)
    \left(\frac{{\rho_\mathrm{ast}}}{2.7 \; \mathrm{g/cm^3}}\right) 
    \left(\frac{{\rho_\mathrm{frag}}}{2.7 \; \mathrm{g/cm^3}}\right)^\frac{1}{2}.
\end{split}
\end{equation}
If the fragment size distribution is instead described by the steeper value of $\alpha = 3.8$, the PR rate becomes:
\begin{equation}
\begin{split}
    \dot{M}_\mathrm{PR, 3.8} &\simeq 2.0 \cdot 10^{8} \; \mathrm{g/s} \; 
    \left(\frac{t_\mathrm{acc}}{10^4 \; \mathrm{yr}}\right)^{-\frac{4}{5}}
    \left(\frac{L_\mathrm{WD}}{0.01\; \mathrm{L_\odot}}\right) 
    \left(\frac{R_\mathrm{max}}{1\; \mathrm{km}}\right)^\frac{4}{5}\\
    & \left(\frac{{R_\mathrm{ast}}}{100\; \mathrm{km}}\right)^3
    \left(\frac{f_\mathrm{bound}}{1}\right)
    \left(\frac{{\rho_\mathrm{ast}}}{2.7 \; \mathrm{g/cm^3}}\right) 
    \left(\frac{{\rho_\mathrm{frag}}}{2.7 \; \mathrm{g/cm^3}}\right)^\frac{4}{5}.
\end{split}
\end{equation}
\subsection{Peak accretion rates by collision-less PR drag}
\begin{figure}
\centering
\includegraphics[width=\hsize]{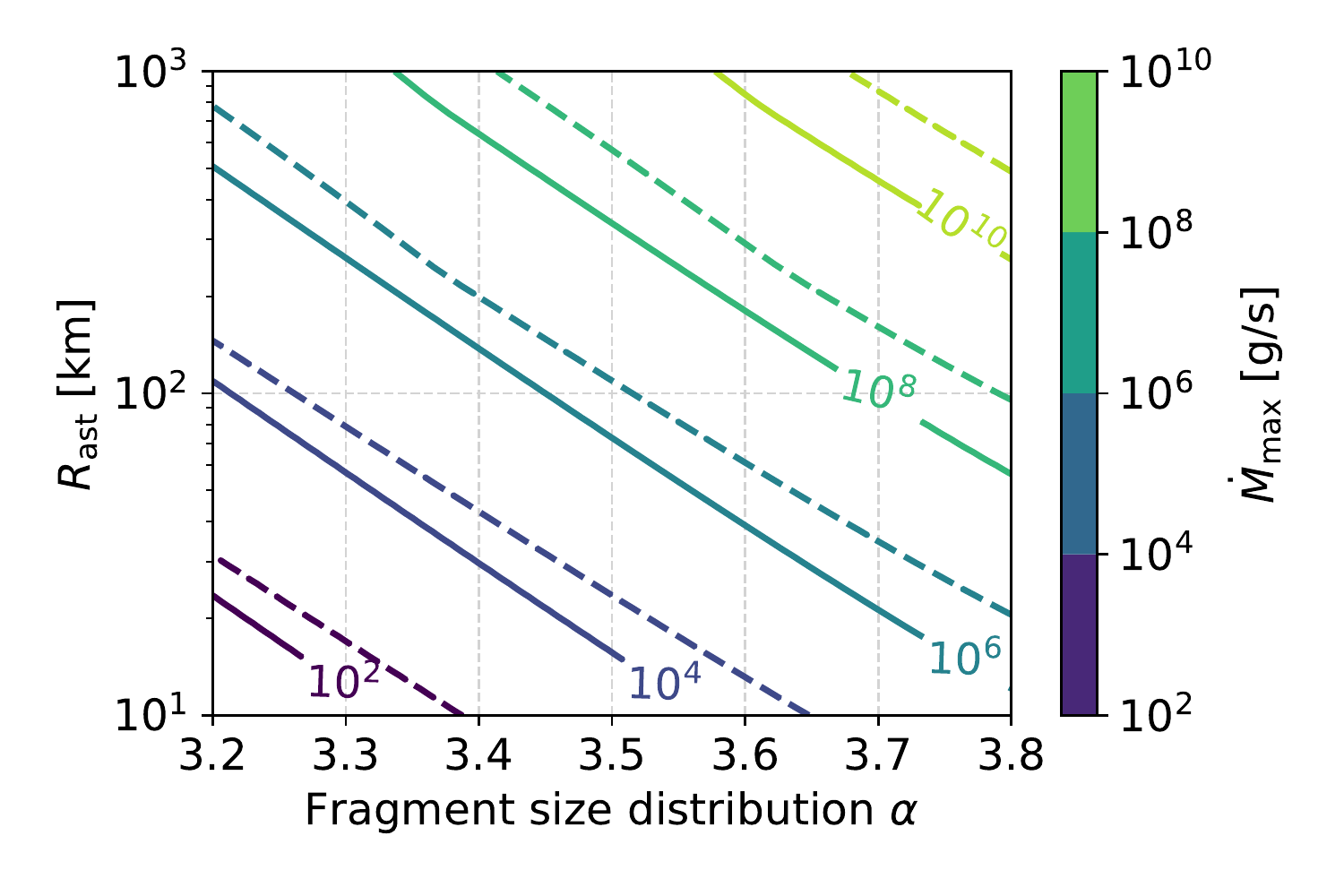} 
\caption{Peak collision-less accretion rates via PR drag onto a warm ($10^4$ K) 0.6 $\mathrm{M_\odot}$ white dwarf as a function of asteroid radius and fragment size distribution. The solid and (dashed) lines correspond to asteroid semi-major axes of 3 and (10 AU). Explaining observed implied accretion rates beyond $10^{9}$ g/s \citep{Farihi2012b} by PR drag alone requires large asteroids ($R_\mathrm{ast}>500$ km) to break up into a steep fragment size distribution ($\alpha > 3.6$). \label{fig:peak_accretion_analytical}} 
\end{figure}
The peak accretion rate by PR drag occurs soon after the asteroid disrupts and the smallest fragments with size $R_\mathrm{blow}$ (Eq. \ref{eq:R_blow}) begin to reach the white dwarf. We evaluate these peak rates in Fig. \ref{fig:peak_accretion_analytical} for a range of asteroid sizes and size distributions. We take a warm white dwarf with temperature $T_\mathrm{WD} = 10^4$ K, which is characterised by both rapid PR-circularisation compared to cooler stars but also a larger blow-out size (Eq. \ref{eq:R_blow}). Of these two, the luminosity effect is more important so the plotted rates can be seen as upper values that decrease slightly for cooler stars. Fig. \ref{fig:peak_accretion_analytical} shows that collision-less accretion via PR drag can at least briefly supply the lowest detectable accretion rates around $10^6$ g/s in the standard case of $\alpha=3.5$. The higher values of observed inferred accretion rates beyond $10^{9}$ g/s \citep{Farihi2012b} are more difficult to explain with PR drag alone and require large asteroids ($R_\mathrm{ast}>500$ km) to break into sufficiently small particles ($\alpha > 3.6$). In addition, the problem of explaining the high accretion rates by PR drag alone is exacerbated for cooler stars with lower luminosities.

In any case, PR drag alone is unable to accrete fragments above $\sim$ 10 cm before the star cools down (see Fig. \ref{fig:M_dot_analytical}). Most of the fragment mass is contained in much larger bodies unless the size distribution is incredibly steep ($\alpha \geq 4$). If we consider the default value of $\alpha =3.5$ and a maximum fragment size of 1 km, the total mass fraction that can be accreted via PR drag is only $(10 \; \mathrm{cm} / R_\mathrm{max})^{1/2} = 0.01$. Because of this accretion inefficiency, the model is in conflict with the higher observed average accretion rates for some DBZ white dwarfs with longer sinking timescales \citep{Farihi2012b} and an additional process is required to explain the observed accretion rates. We will examine how collisions can play a role in increasing accretion efficiency in the next sections.

\section{Stage II\texorpdfstring{\MakeLowercase{a}}{}: Collisions induced by orbital perturbations}\label{sect:perturbations}
We propose the collisional grind-down of larger fragments into dust as a solution to the difficultly in obtaining high accretion rates. In this section, we discuss several processes that naturally lead to the required collisions between fragments soon after the eccentric tidal discs are formed.

\subsection{Differential geodetic precession}\label{sect:diffprec}
Collisions between fragments only occur when their relative orbits are altered sufficiently from their initial trajectories that they begin to overlap. One way in which these changes could be induced, is from the fact that orbits do not precisely follow Newtonian tracks. Instead, their pericentres precess over many orbits according to GR (to lowest order) at a rate ($\dot{\phi}_\mathrm{GR}$) of \citep[i.e.][]{Ragozzine2009, Krishnan2020}:
\begin{equation}\label{eq:dphi_dt}
    \dot{\phi}_\mathrm{GR} = \frac{3}{c^2(1-e^2)} \left(\frac{G^3 M_\mathrm{WD}^3}{a^5}\right)^\frac{1}{2}, 
\end{equation}
which can lead to the build-up of apsidal differences between fragment orbits over time.
In the context of white dwarf pollution, differential apsidal precession was first mentioned by \citet{Debes2012a} and further examined by \citet{Veras2014}, who suggested it can be triggered by orbital differences induced by PR drag. Because the precession rate is highest for the most eccentric orbits and those with short periods, apsidal precession translates orbital differences into angular differences. However, as we discussed in Sect. \ref{sect:stream_morphology}, the tidal breakup process already spreads the fragments along a range of semi-major axes and eccentricities. This means that no additional process is required and that the fragments start differentially precessing as soon as they form. The inner fragments precess more quickly than those on wider orbits, with the timescale for complete differential precession of the inner and outer ring (1 and 2) is given by Eq. \ref{eq:dphi_dt}:
\begin{subequations}
\begin{align}
    \delta \tau_\mathrm{precess} &= \frac{2\pi}{\dot{\phi}_\mathrm{GR,1} - \dot{\phi}_\mathrm{GR,2}} \\
    &\simeq \frac{2\pi c^2 r_\mathrm{B}^3}{9 R_\mathrm{ast}} \left(\frac{a_0}{G^3M_\mathrm{WD}^3}\right)^\frac{1}{2}, \label{eq:dt_precess}
\end{align}
\end{subequations}
in the non-dispersive limit where $R_\mathrm{ast}<<R_\mathrm{crit}$. For partially unbound orbits, the precession timescale is instead given by $2\pi / \dot{\phi}_\mathrm{GR}$.
We visually indicate the differential apsidal precession of an eccentric tidal disc in Fig. \ref{fig:precession_orbits}, where we plot the fragment orbits of a 100 km asteroid that originates from 3 AU (corresponding to the inner boundary of the zone that is not swallowed by the progenitor star during post main-sequence evolution, see \citealt{Mustill2012}) at 0.1 and 1 Myr. While this is less than the timescale of complete orbital precession ($\sim 20 \; \textrm{Myr}$), the highly eccentric nature of the tidal disc means that collisions already occur at small angular differences. Initially, the orbit crossings are restricted to the pericentre and apocentre of the orbit. The relative fragment velocities scale with the angular differences and increase as a function of time. As the angular differences increase, the locations of orbit crossings eventually spread out over a wide range in space.
\begin{figure}
\centering
\includegraphics[width=\hsize]{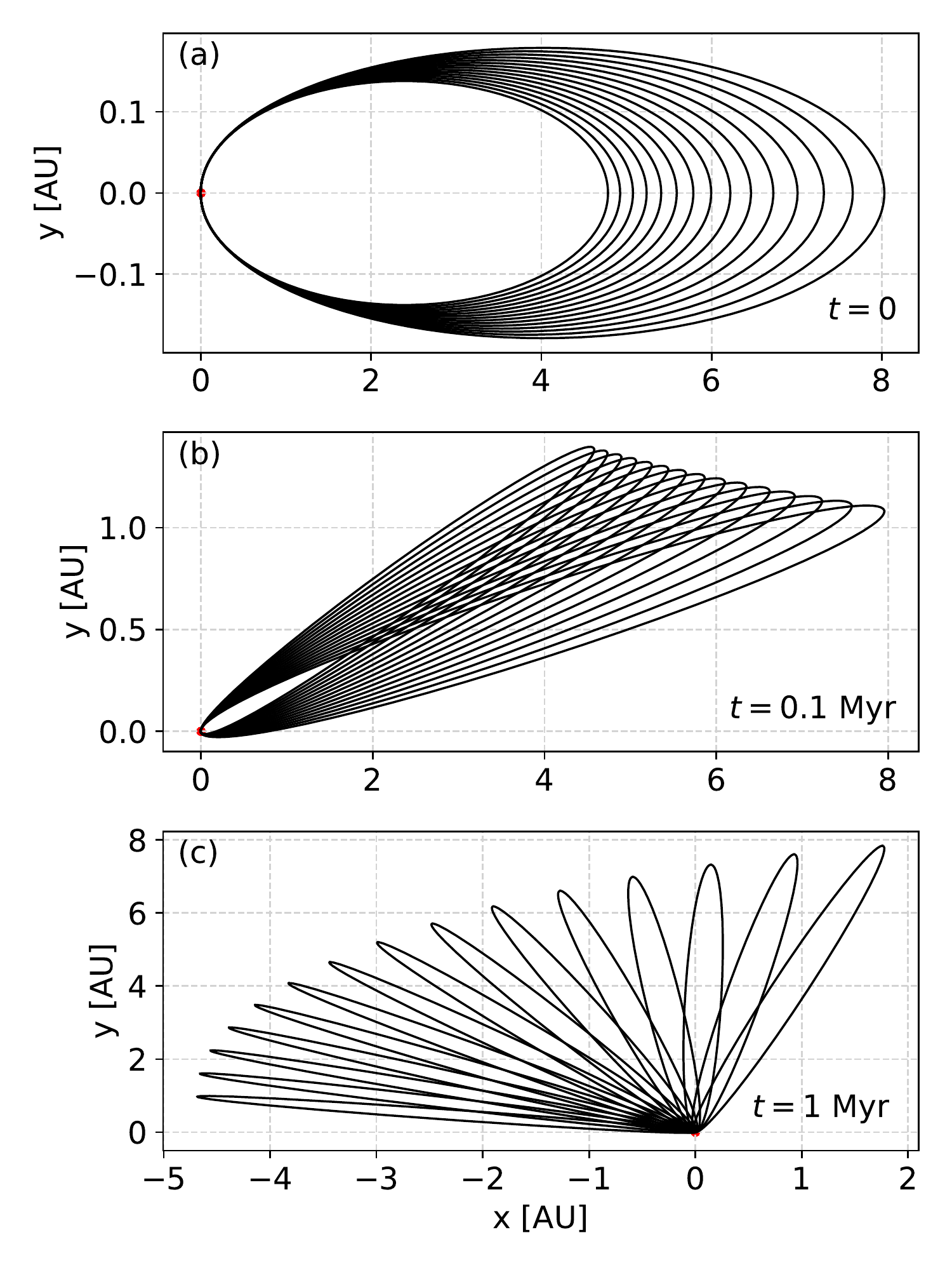} 
\caption{Differential apsidal precession of fragment orbits from a 100 km asteroid originating from 3 AU. The orbits do not cross at $t=0$ (panel a) but angular differences increase as a function of time, and at 0.1 Myr (panel b) the orbits are clearly seen to cross at both pericentre and apocentre. At 1 Myr (panel c), both the relative angles and collision velocities have grown further and the collision locations are spread out over a wider range in space.
\label{fig:precession_orbits}} 
\end{figure}

\subsubsection{Differential precession vs coherent precession}\label{sect:diffprec_2}
Before we proceed with modeling the collisional grind-down induced by differential apsidal precession in the next section, we should note that not all eccentric astrophysical discs are observed to precess at differential rates. For instance, the asymmetric nucleus of the Andromeda galaxy (M31) likely consists of an eccentric disc \citep{Tremaine1995, Lauer2005} that does not show signs of differential precession, even though its stellar ages ($\sim$ Gyr) far exceed the timescale of differential apsidal precession ($\sim$ Myr). We should, therefore, first investigate whether the same processes that act here could cause the orbits of fragments around white dwarfs to precess coherently as well. The process that has been suggested to explain the coherent precession of M31 is a dynamical oscillation of the eccentricities as a result of self-gravity \citep{Madigan2018, Wernke2019}. Physically, any orbits that precess faster than the bulk of the disc are pulled back while orbits that lag behind are pulled forward. This force torques the orbit and changes its angular momentum and eccentricity, generating an oscillation. Based on the calculations of \citet{Madigan2018}, the typical period of these oscillations is the secular timescale $t_\mathrm{sec} = P \left(\frac{M_\mathrm{central}}{M_\mathrm{disc}}\right)$, which is on the order of 100 orbits for M31. In the case of a disruption of an asteroid around a white dwarf, the mass ratio of the central object is much larger, however. Even in the case of a disrupting Earth-mass planet, the mass ratio ($0.6\; \mathrm{M_\odot} / \mathrm{M_\oplus}$) leads to a characteristic scale around $10^5$ orbits, too long to prevent orbit crossings via differential apsidal precession. 

A second option is that differential precession may be inhibited by continual collisions between fragments. If enough collisions occur at small relative angles and velocities, their exchange of angular momentum could prevent further differential precession similar to what happens in eccentric gaseous discs. For this process to be effective, the disc needs to contain a sufficient fraction of small particles to generate a large collisional cross section. In the case of the tidal discs that form from tidal disruptions around white dwarfs, this is unlikely due to the blow-out of the smallest fragments (Sect. \ref{sect:size_dist}) and the typically rapid accretion of slightly larger dust grains. Hence, it seems safe to assume that differential apsidal precession does indeed proceed uninhibited for the larger fragments contained in the tidal debris discs around white dwarfs.

\subsection{Gravitational perturbations by a planet}\label{sect:planet_scattering}
Besides the gravitational perturbations from the central star, gravitational interactions with surrounding planets can also drive collisions between the eccentric fragments. Polluted white dwarfs are thought to be surrounded by whole planetary systems - including potentially gas- and ice giants as well as rocky planets (see \citet{Veras2021b} for a recent review). Because the tidal disc that forms in a disruption event is typically centered around the asteroid's prior orbit, interactions with the planet that scattered it will continue to perturb the surviving fragments \footnote{During the preparation of this manuscript, we became aware of the contemporary manuscript by \citet{Li2021}. These authors have independently come to similar conclusions as presented here, based on numerical simulations of the continued scattering of eccentric fragments.}. 

The effectiveness of this continued scattering likely depends on the geometry of the tidal disc. Any fragments whose orbits either cross the path of the planet or whose apocentre is close to the planet's semi-major axis will continue to be scattered. The width of this chaotic zone ($\delta a_\mathrm{chaos}$ )around the planet has been studied in detail in previous works - albeit with far less eccentric orbits - to be around $\delta a_\mathrm{chaos} = C \; a_\mathrm{pl} \left(\frac{M_\mathrm{pl}}{M_\mathrm{\star}}\right)^\frac{1}{3}$ with constant $1.3 < C < 2$ \citep{Wisdom1980, Duncan1989, Quillen2006, Chiang2009}. If the planet is located at the apocentre of the fragment orbits, this criterion translates to a width of orbit-crossing ($\delta a_\mathrm{i,cross}$):
\begin{subequations}
\begin{equation}
    \delta a_\mathrm{i,cross} \geq \frac{a_\mathrm{pl}}{2}\left[1-C \left(\frac{M_\mathrm{pl}}{M_\mathrm{\star}}\right)^\frac{1}{3}\right],
\end{equation}
\end{subequations}
\begin{figure}
\centering
\includegraphics[width=\hsize]{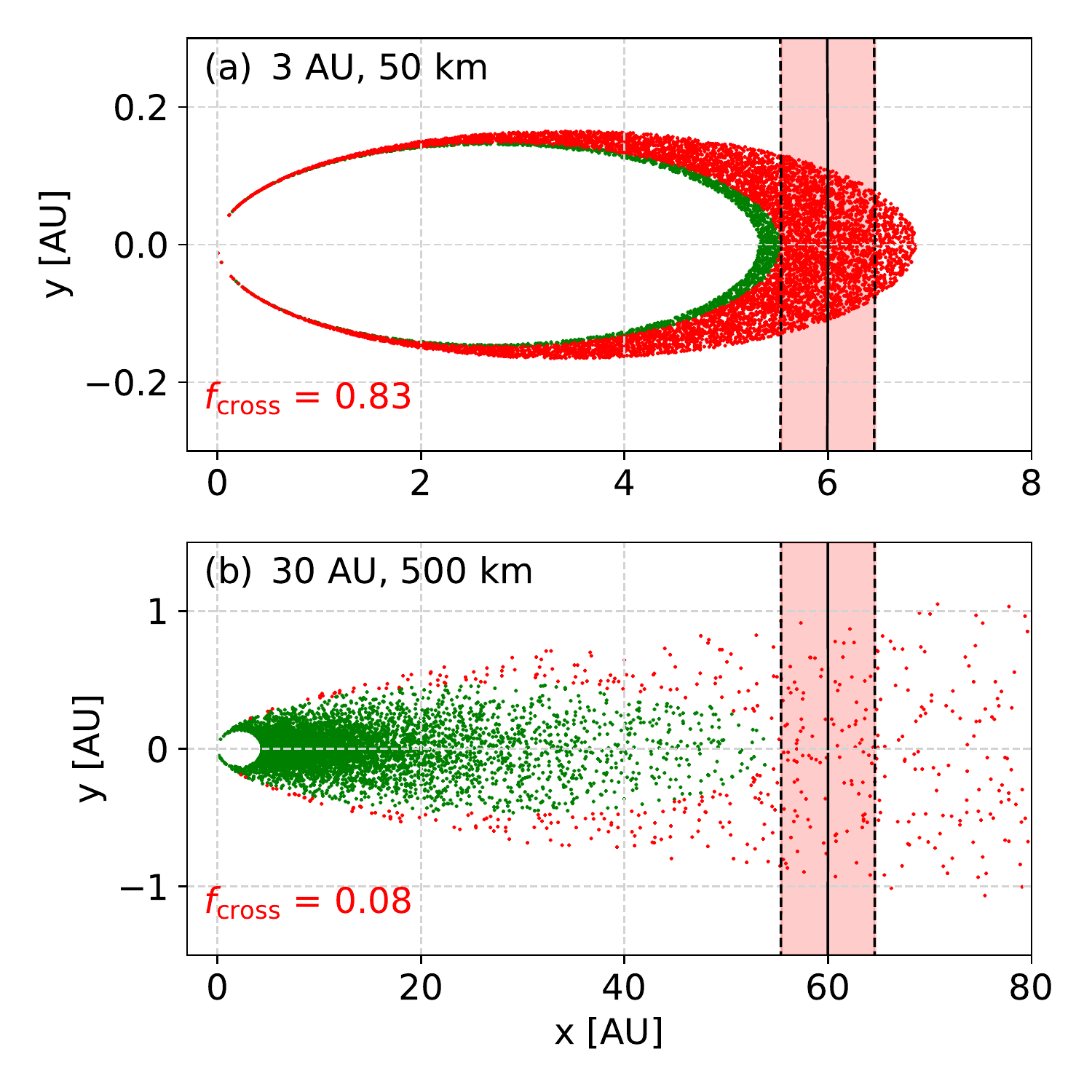} 
\caption{Continued scattering of fragments that cross the chaotic zone of a planet that resides at the apocenter of a scattered asteroid. This figure depicts the eccentric tidal disc, dividing the fragments between those that are likely to be directly re-scattered by the planet (red) and those that now lie outside the planet's chaotic zone (green). The top panel (a) shows that most fragments from a 50 km asteroid from 3 AU are susceptible to scattering (fraction $f_\mathrm{cross}$). Fragments from a larger 500 km asteroid that originates from 30 AU (panel b) are spread bimodally (see Sect. \ref{sect:disruption}) and are mostly safe from further scattering.
\label{fig:orbits_scattering}} 
\end{figure}
where we use that the eccentricities of fragments in these tidal discs are close to unity. We visualise the range of this scattering for 100 and 500 km asteroids with semi-major axes of 3 and 30 AU in fig. \ref{fig:orbits_scattering}. If the asteroid is small and originates from the inner zone, it will form a narrow orbital band and most of its fragments lie within the chaotic zone. Larger asteroids or those from further out are less likely to be perturbed directly, as the fragment orbits begin to follow a bimodal bound-unbound distribution, with up to half of the fragments tightly bound to the star with orbits that are safe from direct scattering. Nevertheless, slower perturbations of these inner orbits are also possible around interior mean-motion resonances or via secular perturbations if the planetary orbit is not entirely circular, as discussed by \citet{Veras2021a}.

Over time, these orbital perturbations will lead to collisions between fragments and facilitate faster collisional grind-down. In addition, some fragments will be scattered out of the system while others are scattered to bound orbits with reduced pericentre distances. We suggest that fragment scattering in this way can provide a separate avenue for white dwarf pollution. Instead of having to lose angular momentum through stellar light, some fragments will collide with the white dwarf directly. Others are scattered into the sublimation zone, where the vapor pressure of their material becomes significant and they and disintegrate over several orbits.

\subsection{Drag-assisted circularisation via pre-existing material}
Thirdly, fragments can change their orbits by interacting with pre-existing gaseous or dusty material inside the Roche radius \citep{Grishin2019, Oconnor2020, Malamud2021}, such as has been observed around a substantial minority of systems \citep{Rocchetto2015, Farihi2016, Wilson2019}. The near-unity eccentricities of fragments mean that their pericentre velocities are typically on the order of several 100 km/s, yielding an extremely violent interaction with any dust or gas that is encountered. In a detailed study, \citet{Malamud2021} showed that this interaction can significantly alter fragment orbits. Firstly, the fragments lose kinetic energy and become more tightly bound to the star. Interestingly, the orbital contraction already becomes significant when the mass of the central compact disc is several orders of magnitude below that of the tidally disrupted asteroid. The smallest fragments circularise the fastest, typically within a few orbits. Larger fragments require additional passages through the disc, causing increased orbital differences that again accelerate the onset of orbit crossings via differential apsidal precession. In addition, regardless of whether the pre-existing disc contains dust or gas, each fragment passage though the central compact disc erodes away a mass similar to the mass that is encountered. Complete circularisation of a fragment requires colliding with a similar amount of mass as the fragment itself and therefore also leads to its complete disintegration. In the presence of such a massive central compact disc, fragments can accrete onto the white dwarf without the necessity for collisional grind-down.

\subsection{Orbital changes due to the Yarkovski effect} 
Finally, larger fragments ($\geq$ 0.1-100 m) may either gain or lose angular momentum over time due to the Yarkovski effect \citep{Bottke2006, Veras2015a,Veras2015b}. The idea is that while a fragment orbits the white dwarf, its side that faces the star is more strongly irradiated. Subsequent re-emission occurs with a time lag that, coupled with a rotation, leads to either an accelerating or braking term. If the fragment spins sufficiently quickly, its temperature gradient smooths out and the term disappears. While the Yarkovski effect likely dominates over PR drag in most larger fragments, its derived terms are highly dependent on poorly constrained fragment parameters like the spin period (see \citealt{Veras2015a,Veras2019}). While a first attempt at constraining the spin distribution was made by \citet{Malamud2020b}, current simulations are not yet able to resolve them for physical fragment sizes. These characteristics of the Yarkovski effect currently make a useful inclusion in a collisional model unfeasible. However, it is clear that by increasing or decreasing the orbital momentum of different fragments, the Yarkovski effect will both facilitate collisions directly and induce orbital differences that accelerate the process of differential apsidal precession.

\section{Stage II\texorpdfstring{\MakeLowercase{b}}{}: Calculation of eccentric collisional grind-down}\label{sect:collisional_model}
In this section, we formulate a crude but quantitative calculation of collisional grind-down within the highly eccentric tidal discs that form after asteroids tidally disrupt around a white dwarf. Our model is based on the angular differences that are induced by differential apsidal precession (Sect. \ref{sect:diffprec}), which in turn originate from the orbital spread imparted at the moment of tidal breakup (Sect. \ref{sect:stream_morphology}). As discussed in the previous section, there are many additional mechanisms that also drive orbital differences between fragments. We take differential precession as the sole perturbing process here to make the calculation tractable and because it is universally applicable to tidal discs around white dwarfs, that all start with the required orbital spread.

\subsection{Numerical setup}
\begin{figure}
\centering
\includegraphics[width=\hsize]{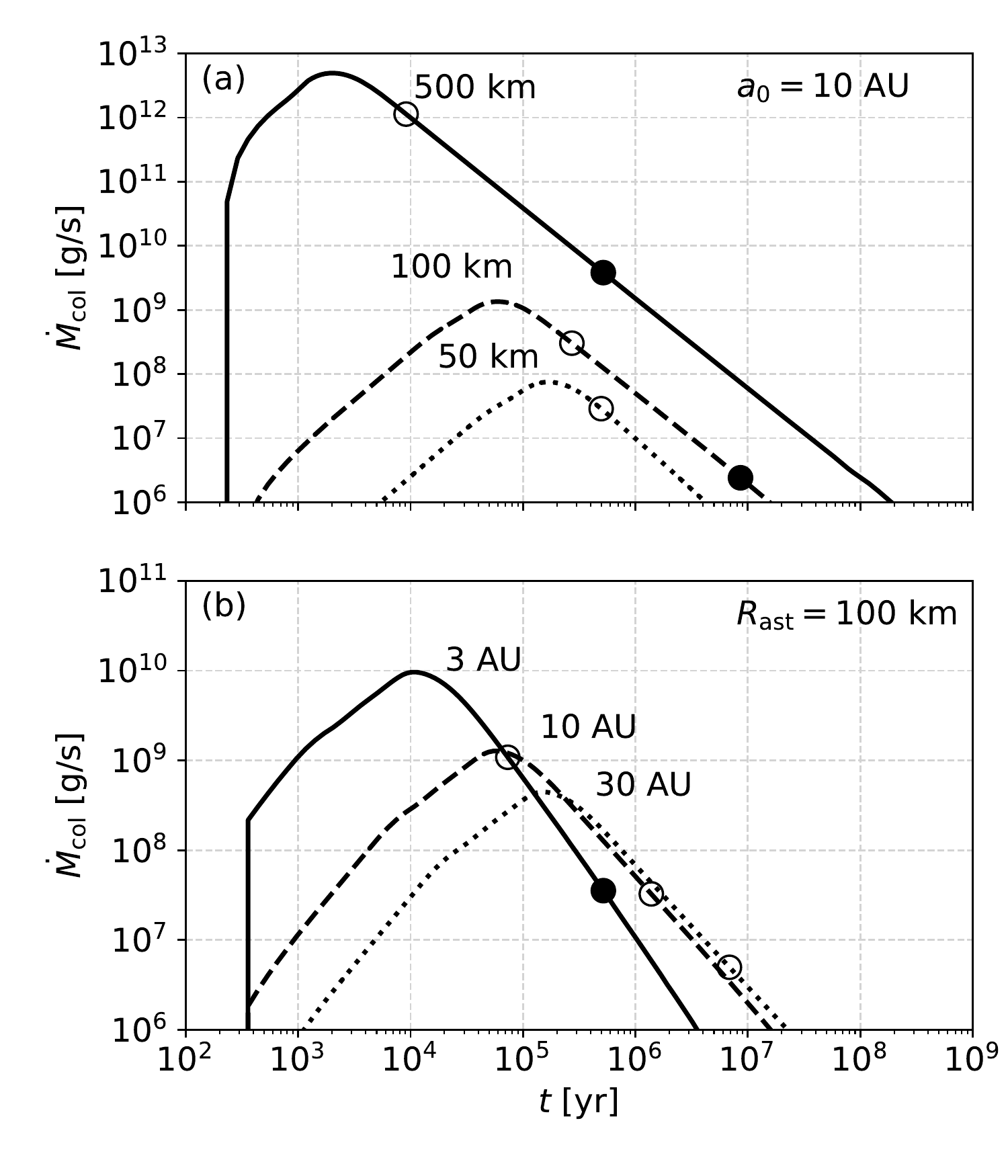} 
\caption{Time evolution of the collision rate ($\dot{M}_\mathrm{col}$) from our model of eccentric collisional grind-down induced by differential apsidal precession (see text for details). In the top panel (a), we take asteroids from 10 AU and vary their size between 50-500 km. In the bottom panel (b), we take 100 km asteroids and vary their initial semi-major axis between 3-30 AU. The open and filled dots indicate the points where a total of 50\% and 90\% of the fragment mass has catastrophically collided, respectively. Larger asteroids on tighter orbits disrupt to form tidal discs whose fragments collide within the shortest period of time. \label{fig:example_runs}} 
\end{figure}
We opt for a simple computational approach where we divide the fragments into a 2D grid along semi-major axis and fragment size and model the collisions with a particle-in-a-box method. The fragments are assumed to be spread evenly across energy bins, with semi-major axes and eccentricities that lie between the bounds specified by Eq. \ref{eq:a_new_1}-\ref{eq:e_orbits}. Their initial sizes are assumed to follow the collisionally evolved distribution of Eq. \ref{eq:dm_dR_dist} with $\alpha =3.5$ as a typical value, meaning that most of the mass is contained in the larger fragments, whereas their surface area is dominated by the smaller fragments. As described in Sect. \ref{sect:diffprec}, the orbits precess at different rates depending on their eccentricities and semi-major axes (Eq. \ref{eq:dphi_dt}), leading to orbit crossings. We estimate the rate of collisions in a projected 2D plane, where the collision points can be identified at radius $r_\mathrm{col}$ and true anomaly $\theta_\mathrm{col}$ for any two bins with indices $i,j$ from the criterion that $r_\mathrm{i} (\theta_\mathrm{col}, \phi_\mathrm{i}) = r_\mathrm{j} (\theta_\mathrm{col}, \phi_\mathrm{j}) = r_\mathrm{col}$, where the velocity $v_\mathrm{i}$ and distance $r_\mathrm{i}$ are given by the standard equation of the ellipse in polar form:
\begin{subequations}
\begin{align}
    r_\mathrm{i} (\theta) &= \frac{a_\mathrm{i}(1-e_\mathrm{i}^2)}{1-e_\mathrm{i}\mathrm{cos(\theta_\mathrm{i} - \phi_\mathrm{i})}}, \label{eq:r_theta}\\
    v_\mathrm{i}(r) &= \left(GM_\mathrm{WD}\left(\frac{2}{r_\mathrm{i}}-\frac{1}{a_\mathrm{i}}\right)\right)^\frac{1}{2}.
\end{align}
\end{subequations}
While all collisions exchange some angular momentum, only those that are sufficiently violent lead to the (partial) destruction of fragments. Accounting for the angular momentum changes in sub-catastrophic collisions is not possible with our method due to the large number of additional spacial bins it would create. Hence, we do not include these less violent collisions in our calculation and only focus on the collisional grind-down from catastrophic collisions. The required specific energy for catastrophic fragmentation $Q = \frac{1}{2} (\vec{v_\mathrm{i}} - \vec{v_\mathrm{j}})^2 \frac{m_\mathrm{j}}{m_\mathrm{i}}$ is known as the dispersion threshold and can be estimated as a scaling relation \citep{Durda1998, Benz1999}:
\begin{equation}\label{eq:Q_star}
    Q^\star = Q_\mathrm{a} R^{-a} + Q_\mathrm{b} R^b,
\end{equation}
where the first term accounts for the material strength of a fragment and the second term corresponds to the gravitational binding energy that has to be overcome. We take the constants $a=0.3, \; b=1.5$ and $Q_\mathrm{a} = 6.2\cdot 10^7 \; \mathrm{erg/g}, \; Q_\mathrm{b} = 5.6 \cdot 10^{-2} \; \mathrm{erg/g}$ from \citet{Lohne2008} and \citet{Wyatt2011}. In our simulations, fragments of different sizes that reside in the same orbital bin experience the same collisional velocities when they cross paths with fragments in other bins. This means that Eq. \ref{eq:Q_star} also directly specifies the minimum fragment size $R_\mathrm{j, crit}$ that can catastrophically collide with a fragment of size $R_\mathrm{i}$, depending on its collisional velocity:
\begin{equation}\label{eq;R_crit}
    R_\mathrm{j, crit} = \left(\frac{2Q^{\star}}{v_\mathrm{col}^2}\right)^\frac{1}{3} R_\mathrm{i}.
\end{equation}
Generally, only fragment pairs with size ratio's below two orders of magnitude are found to collide catastrophically.

The collision rate $P_\mathrm{ij}$ of two bins with fragment sizes $R_\mathrm{i}, R_\mathrm{j}$ and $N_\mathrm{i}, N_\mathrm{j}$ fragments can be estimated from a standard particle-in-a-box approach as:
\begin{equation}
    P_\mathrm{ij} = \frac{<T_\mathrm{box, i}>}{T_\mathrm{orb, i}} \frac{<T_\mathrm{box, j}>}{T_\mathrm{orb, j}} \frac{N_\mathrm{i} N_\mathrm{j} \sigma_\mathrm{ij} v_\mathrm{col, ij}}{V_\mathrm{box}} \label{eq:Pcol_1},
\end{equation}
where $\sigma_\mathrm{ij} = \pi (R_\mathrm{i}+R_\mathrm{j})^2$ is the collision cross section, $V_\mathrm{box}$ is the volume of the collision box and $<T_\mathrm{box, i}>,<T_\mathrm{box, j}>$ are the average periods of time that fragments of bins $i$ and $j$ spend there. We approximate these collision volumes as locally straight boxes with volume $V_\mathrm{box} \simeq l_\mathrm{i} l_\mathrm{j} H \mathrm{sin}(\alpha_\mathrm{ij})$ where $\alpha_\mathrm{ij} = |\vec{v_\mathrm{i}} \times \vec{v_\mathrm{j}}| / (|\vec{v_\mathrm{i}}||\vec{v_\mathrm{j}}|)$ is the angle between the orbits and $H$ is the local height of the tidal disc. We estimate this height based on a typical orbital inclination $\bar{i}$ as $H_\mathrm{i} = 2\bar{i} r_\mathrm{col}=2R_\mathrm{ast} r_\mathrm{col}/r_\mathrm{B}$ (see Sect. \ref{sect:stream_morphology}), assuming that the inclination remains constant over time. The widths $l_\mathrm{i}$ and $l_\mathrm{j}$ cancel from the time spent in the box <$T_\mathrm{box, i}> = l_\mathrm{i}/|\vec{v_\mathrm{i}}|$, which gives:
\begin{equation}
    P_\mathrm{ij} =  \frac{\pi N_\mathrm{i} N_\mathrm{j} (R_\mathrm{i} + R_\mathrm{j})^2 |\vec{v_\mathrm{i}} - \vec{v_\mathrm{j}}|}{T_\mathrm{orb, i} T_\mathrm{orb, j} H |\vec{v_\mathrm{i}} \times \vec{v_\mathrm{j}}|} \label{eq:Pcol_2}.
\end{equation}
The total collision rate of bin i is set by the binned sum over $P_\mathrm{i,j}$:
\begin{equation}\label{eq:Pi_sum}
    P_\mathrm{i} = \sum_{j_\mathrm{ crit,min}}^{j_\mathrm{crit,max}}{P_\mathrm{ij}},
\end{equation}
which we evaluate numerically. Because we only track catastrophic collisions in this scheme, we register the first catastrophic encounter for any fragment and remove the pair involved in the collision for the rest of the simulation. Although this approach does not incorporate grind-down that can result from second generation fragments, larger fragments generally require more time to catastrophically collide so our approach can still provide order-of-magnitude estimates. More importantly, even this crude model can illuminate important trends and biases that are inherent to the collisional phase, which we will describe in the next subsections.

\subsection{General trends in eccentric grind-down rates}
\begin{figure}
\centering
\includegraphics[width=\hsize]{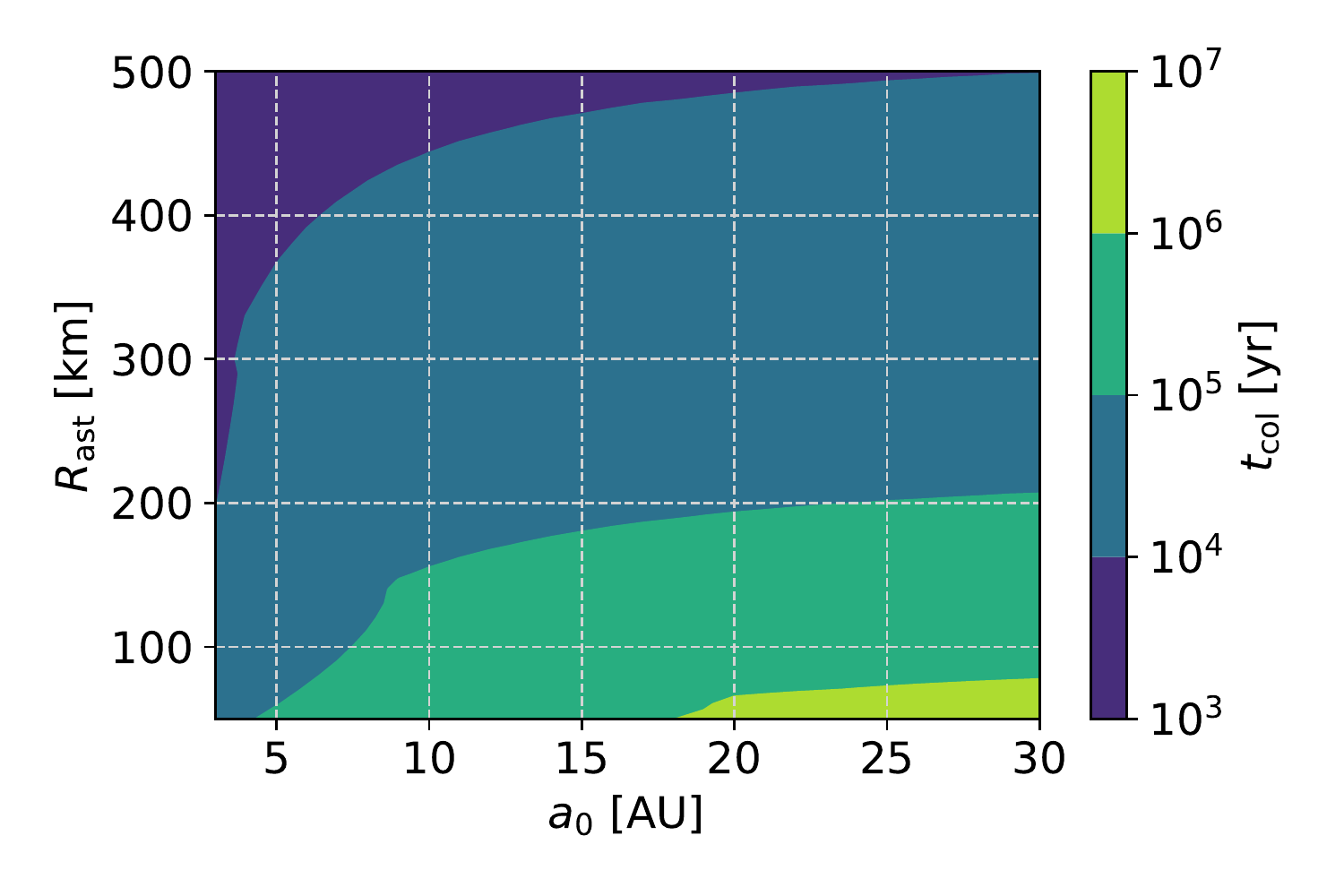} 
\caption{Timescales for the collisional grind-down of fragments in eccentric tidal discs ($t_\mathrm{col}$), calculated with our numerical model based on differential apsidal precession (see text for details). The grind-down timescale is defined as the time required for half of the mass to catastrophically collide. Fragments within tidal discs that form from large asteroid progenitors on tight orbits grind down within the shortest period of time.
\label{fig:t_col}} 
\end{figure}
We explore the most important trends of eccentric fragment collisions in Fig. \ref{fig:example_runs} where we simulate the grind-down of fragments from a range of asteroid progenitors. In the top panel (a), we take asteroids with fixed semi-major axes of 10 AU but with sizes between 50-500 km. In the bottom panel (b), we instead take fixed asteroid sizes of 100 km but vary their semi-major axis between 3-30 AU. The first thing to note is that their rates of grind-down follow similar temporal shapes. Initially, no fragments cross paths and the rate collisions begins at zero. As the fragment orbits continue to precess at different rates, some orbits begin to cross but the relative velocities are low and only similar-sized fragments catastrophically collide. After $10^2 - 10^4$ years, these relative velocities have increased sufficiently that catastrophic collisions occur for a wide range of relative fragment sizes and the collision rate shoots up. When around half of the fragment mass has collided (open dots), the rate decreases from its peak value and continues to drop as fewer and fewer intact fragments remain.

The most important variable in the rate of grind-down is the size of the asteroid progenitor. Larger asteroids provide more fragments that can collide (factor of $R_\mathrm{ast}^3$), which additionally means that each fragment has more other fragments that it can collide with (also a factor of $R_\mathrm{ast}^3$). Together, this predicts a scaling of the grind-down rate of $\dot{M}_\mathrm{col} \propto R_\mathrm{ast}^6$, similar to what we find in our simulations. In Fig. \ref{fig:example_runs}a, we find that the peak accretion rate increases from $\sim 10^8$ g/s for a 50 km asteroid from 10 AU to $\sim 10^{13}$ g/s for a 500 km asteroid with the same semi-major axis. This slightly flatter scaling ($\propto R_\mathrm{ast}^5$ instead of $\propto R_\mathrm{ast}^6$) is due to the more disperse disc that forms when a larger asteroid disrupts (see Figs. \ref{fig:disruption_width} and \ref{fig:new_orbits}). Some fragments of the 500 km progenitor are placed on unbound orbits and are ejected form the system, while others are just placed on wider orbits that take longer to collide. This general inverse scaling of the rate of grind-down with the semi-major axis of the asteroid is also shown in Fig. \ref{fig:example_runs}b, where a 100 km asteroid from 3 AU is found to have a peak collision rate that exceeds the collision rate of an asteroid from 30 AU by around an order of magnitude.

\subsection{Event lifetimes and peak accretion rates}
In order to compare the results of our simulations to observationally inferred accretion event lifetimes, we simulate a grid of 400 tidal discs corresponding to asteroid sizes between 50-500 km that originate between 3-30 AU. Our results are shown in Fig. \ref{fig:t_col}, where we indicate the time required to grind down half of the fragment mass (labeled $t_\mathrm{col}$). As explained in the previous subsection, our model yields a large variation in collisional timescales based on the asteroid size and its semi-major axis. Rather than try to produce an exact fit to the DAZ population from a crude model, we investigate the observability trends based on these parameters. The shortest grind-down timescales that we find are on the order of $10^3$ years for asteroids that are several hundreds of kilometers in size. When the asteroid size is reduced to 50 km, the timescale increases by orders of magnitude to $10^6-10^7$ yr, depending on the orbital separation of the asteroid. From an observational perspective, typical accretion lifetimes of material around white dwarfs can most directly be inferred from differences in detection rates of DAZ and DBZ stars. In the pioneering study by \citet{Girven2012}, the typical timescales are inferred from the difference between DAZ and DBZ pollution rates to be between $10^4-10^6$ yr. More recently, this analysis was revisited by \citet{Cunningham2021} with updated photospheric modeling, which yielded an order of magnitude longer timescales between $10^5-10^7$ yr. A different approach was taken by \citet{Harrison2021}, who estimated typical accretion event lifetimes around $10^7$ yr based on a Bayesian analysis of the photospheric composition. When compared to our simulation results, lifetimes around $10^4-10^6$ yr match our computed grind-down timescales for 100-400 km asteroids, whereas we only find longer lifetimes of $10^7$ yr when the asteroids are smaller than 50 km.

\begin{figure}
\centering
\includegraphics[width=\hsize]{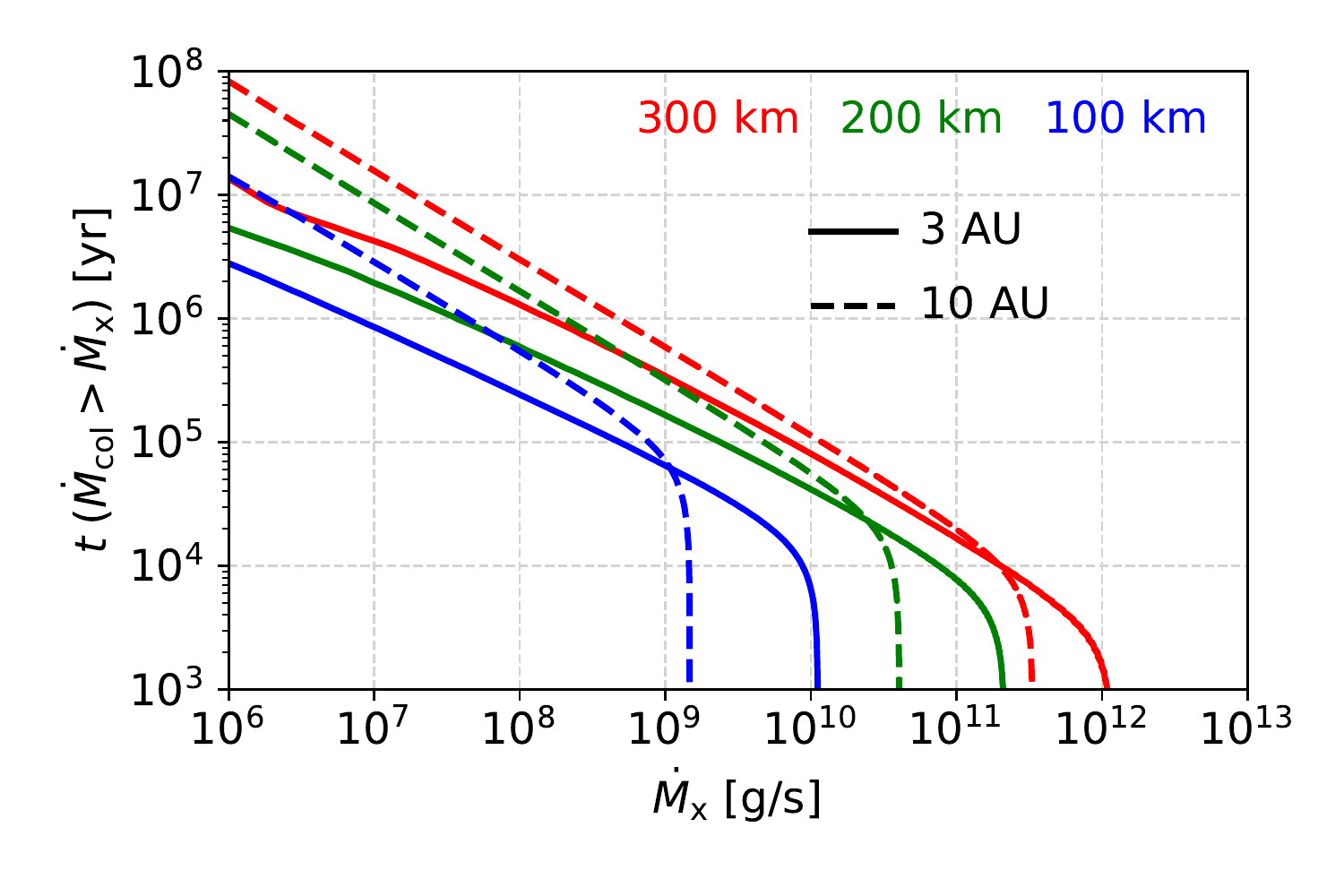} 
\caption{Cumulative time where the rate of dust production from collisional grind-down lies above a certain rate $t(\dot{M}_\mathrm{col}>\dot{M}_\mathrm{x})$. The values correspond to calculations with our collisional model (see text for details). Larger asteroids produce more fragments, leading to more rapid grind-down and higher peak collision rates. Asteroids that originate from the inner zone of planetary systems (solid lines, 3 AU) generate higher peak collision rates compared to those on wider orbits (dashed lines, 10 AU) but they spend less time accreting at more moderate rates, making their pollution less likely to be observed if the subsequent accretion of small dust is sufficiently rapid.
\label{fig:cumulative_single}} 
\end{figure}
The highest inferred on-going accretion rate to date is $10^{9.3}$ g/s for the most metal-rich DAZ \citep{Farihi2012b}. If convective overshoot is accounted for, the inferred rates could increase by another order of magnitude \citep{Cunningham2019}. In Fig. \ref{fig:cumulative_single}, we examine whether such high accretion rates can be generated by the grind-down of asteroids with sizes between 100-300 km that originate from either 3 or 10 AU. The figure shows that this is indeed the case when any produced dust is rapidly accreted, as all lines show at least some period of grind-down rate beyond $10^{9.5}$ g/s. Within the context of our model, grind-down rates beyond $10^{11}$ g/s require asteroids with sizes that exceed 200 km.

The most extreme outliers of the inferred accretion rates have only been observed for DBZ stars. \citet{Farihi2012b} identified six objects with inferred accretion rates above $10^{10}$ g/s, with the highest observation implying a record rate of $10^{11.5}$ g/s. Due to their longer sinking timescales, these are not observations of on-going accretion rates but represent an average over a longer time period. Based on their findings, \citet{Farihi2012b} suggest that short periods of violent accretion that take between 10-100 years must occur at times to explain the difference between DAZ and DBZ upper values. Within the context of our model, the dichotomy between the DAZ and DBZ upper values arises naturally. More massive asteroids produce more fragments that each collide within a shorter period of time, causing the accretion of the biggest asteroids to occur in short and intense bursts, provided that the accretion of the produced dust is sufficiently fast. In contrast, the grind-down of fragments that originate from smaller asteroids occurs over longer periods of time, making them more commonly detected in DAZ pollution, which measures on-going accretion.

\subsection{Monte-Carlo analysis of accretion rate distribution}\label{sect:monte_carlo}
If the dust that is produced in the collisional grind-down of larger fragments is accreted quickly, the collision rate directly translates into an accretion rate onto the star. In this subsection, we explore the trends and biases that this would imply for observations of polluted white dwarfs. We perform Monte-Carlo simulations for a few different asteroid populations, where we scatter and tidally disrupt a total mass of 2.4 $\cdot10^{24}$ g in asteroids over a period of 1 Gyr, which is the equivalence of the mass contained in the Solar System's main belt \citep{Pitjeva2018}. While this amount may seem excessive, it is also equivalent to only 2\% of the Kuiper belt \citep{Fraser2014} or 1\% of the mass of the debris disc around $\zeta$ Leporis \citep{Chen2001}. Spread over 1 Gyr, it amounts to average accretion rate between $3.9-7.8 \cdot 10^7$ g/s.

\begin{figure}
\centering
\includegraphics[width=\hsize]{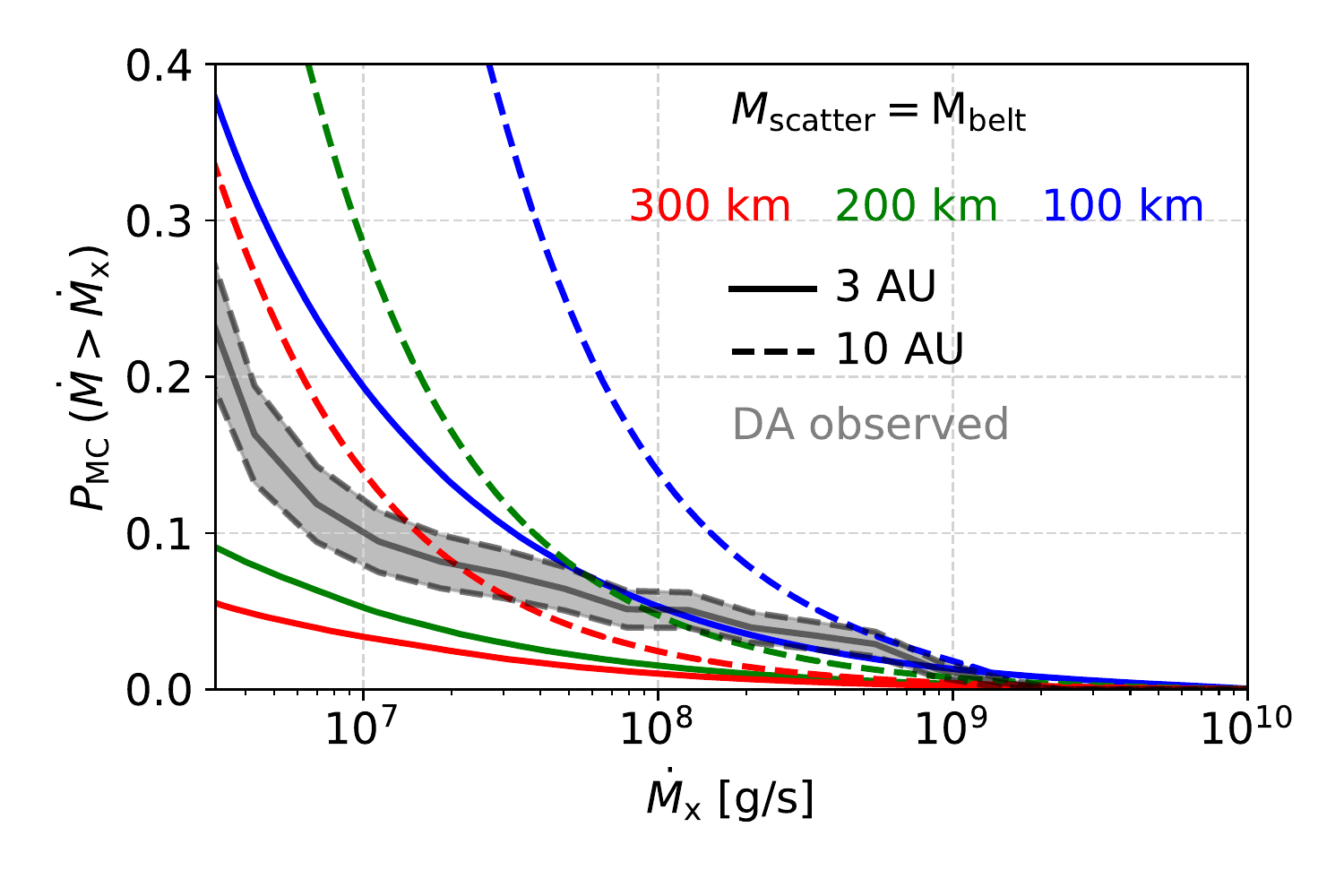} 
\caption{Cumulative probability distribution of grind-down rates (colored curves) compared to an unbiased young DAZ sample of inferred accretion rates (grey, compiled by \citealt{Wyatt2014}), plotted with a 1 $\sigma$ uncertainty band. The grind-down curves correspond to Monte-Carlo simulations of our eccentric collision model (see text for details). In this plot, we examine systems where the equivalence of the main belt ($1.2-2.4\cdot10^{24}$ g) is scattered in over a period of 1 Gyr. The colored lines correspond to mono-size asteroid distributions of 100 km (blue), 200 km (green) or 300 km (red). The asteroids have semi-major axes of 3 AU (solid) or 10 AU (dashed).
\label{fig:cumulative_P}} 
\end{figure}
We contrast the statistical distribution of our calculated grind-down rates with the DAZ sample of \citet{Koester2006} and \citet{Zuckerman2003}, which was compiled and used for the same purpose by \citet{Wyatt2014}. This sample is specifically selected for this purpose because the stars it contains are randomly chosen based on being nearby and bright, and they are not biased in terms of the presence or absence of metals. The sample contains a total of 534 DAZ systems from a Keck and SPY survey, with 38 CA detections and 298 upper limits on CA. We only use the 467 systems that have estimated sinking times less than $10^4$ yr (from the estimate of \citealt{Koester2009}), so that the sample reflects on-going accretion rather than an average over past accretion events. In order to make a uniform comparison across the sample, we follow \citet{Wyatt2014} and calculate the total accretion rate based on the Ca accretion rate with a scaling that matches bulk Earth, noting that true accretion rates may vary from this assumption, particularly in Ca-rich bodies, by orders of magnitude. Using the sinking times from \citet{Koester2009}, the measurements imply accretion rates between $10^{6.1} -10^{9.3}$ g/s. Identical to the methodology of \citet{Wyatt2014}, the cumulative probabilities are calculated from the sub-sample of systems $N_\mathrm{ss} (\dot{M}_\mathrm{x})$ where accretion could have been detected at that level. The 1 $\sigma$ errors are calculated from binomial statistics \citep{Gehrels1986}.

We plot the DAZ sample together with the results of our Monte-Carlo simulations in Fig. \ref{fig:cumulative_P}. The model presented here predicts accretion rates based on two parameters, namely the asteroid size and its initial semi-major axis. As explained in the previous subsection, larger asteroids produce more fragments that each collide within shorter periods of time, and hence appear as short but intense bursts of accretion. This can be compared to the slower, steadier accretion from smaller asteroids (see Fig. \ref{fig:example_runs}). When the same total mass in asteroids is scattered towards the star, a population containing larger asteroids leads to a steep cumulative distribution of the accretion rate, with few white dwarfs found accreting at the highest rates, whilst a population of smaller asteroids leads to a flatter distribution with more white dwarfs found at typical accretion rates but lower peak values. These trends are visualised in Fig. \ref{fig:cumulative_P} from the comparison between curves with the same line style (semi-major axis) but different colors (size). Each of the curves within the same panel represents the same total mass in asteroids scattered in, with the red curves representing accretion of the biggest asteroids (300 km) and the blue corresponding to the smallest asteroids (100 km). While the bigger asteroids have higher peak accretion rates (see also Fig. \ref{fig:cumulative_single}), their cumulative probabilities of accretion at detectable levels ($\gtrsim 10^7$ g/s) are as much as an order of magnitude lower.

The observability bias based on semi-major axis proceeds along the same lines. Asteroids with wider orbits prior to their tidal disruption release their fragments along more spread-out discs. As a result, their fragments each take longer to complete an orbit and require more time to catastrophically collide. This means that the fragments from wide asteroid progenitors spend more time producing dust at lower rates, at the cost of having a reduced peak value (see Fig. \ref{fig:cumulative_single}). In the Monte-Carlo simulations of Fig. \ref{fig:cumulative_P}, we visualise the difference with solid and the dashed curves of the same color, which correspond to the accretion of identical asteroids with semi-major axes of 3 AU and 10 AU, respectively. The effect of orbital separation exposes a second bias towards detecting on-going accretion at moderate values ($\lesssim 10^9$ g/s) by asteroids that originate from wider orbits. 

\section{Stage III: Infrared emission from accreting fragments or dust}\label{sect:IR_Excess}
A substantial minority of accreting white dwarfs exhibit detectable infrared excesses \citep{Rocchetto2015, Farihi2016, Wilson2019}. While the simplest canonical model computes emission from opaque dust on circular orbits \citep{Jura2003}, elliptical discs around white dwarfs have been modeled in response to observational clues to account for the reduced infrared excess around young stars \citep{Dennihy2017} and also to explain the variability of the emission \citet{Nixon2020}. The elliptical nature of the material is furthermore suggested by Doppler tomography of gas near the Roche radius \citep{Manser2016a, Steele2021}. From a physical perspective, it is similarly clear that the fragments are indeed expected to be released on orbits that are initially highly eccentric (see Sect. \ref{sect:stream_morphology}) and then slowly circularise over time. In our modeling, we compute the emission that is generated during the accretion of dust from such a highly eccentric structure. We do not follow the typical assumption that the disc is opaque to stellar light. Instead, we compute the distribution of dust and its re-emission under the a-priori assumption of an optically thin environment, the validity of which we then discuss after we calculate the radial optical depth.

With our model, we show that that typically inferred accretion rates can indeed produce detectable infrared excesses when grains circularise in an optically thin disc via PR drag alone, but that the disc only remains optically thin if the orbital inclinations of the dust are increased over time. We then evaluate the results within our three-stage model and discuss how the disc geometry and circularisation speed can effect the produced infrared excesses.

\begin{figure*}
\centering
\includegraphics[width=\hsize]{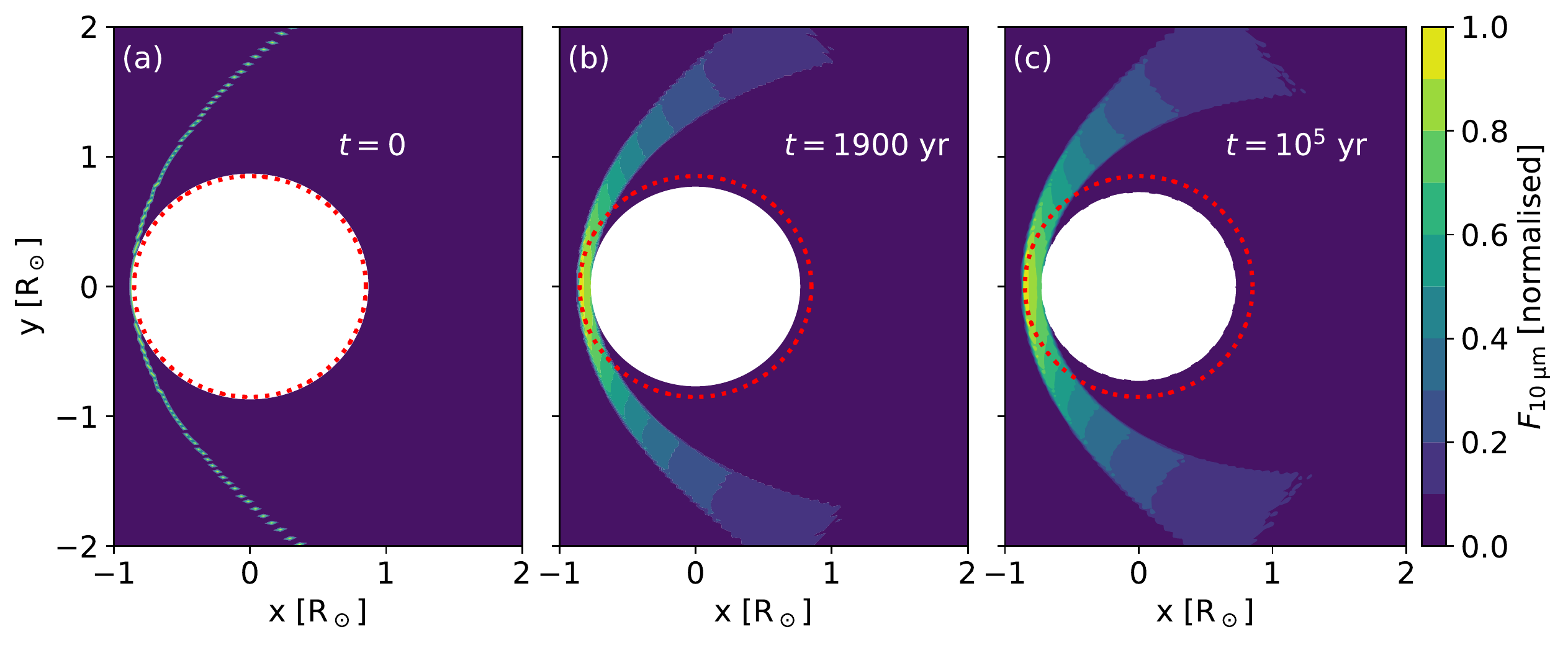} 
\caption{Simulated emission per unit surface area at 10 $\mathrm{\mu m}$ from fragments accreting onto G29-38 ($0.59 \; \mathrm{M_\odot}$, 11240 K, 17.5 pc) via collision-less PR drag, computed assuming optically thin properties (See Sect. \ref{sect:pr_drag} and Sect. \ref{sect:IR_excess_av} for details on PR drag and this emission). The three panels correspond to snapshots at different times after the tidal disruption event of an asteroid from 3 AU and the colors are normalised individually per panel. Just after the tidal disruption event (a), the fragments are still highly eccentric and their emission is minimal. The smallest fragments accrete just after 1900 yr (b), when their more contracted orbits begin to generate far more emission. The final panel corresponds to $10^5$ yr, when all fragments smaller than $100 \; \mathrm{\mu m}$ have already reached the star. The disc contains an elliptical inner gap because fragments are not yet fully circularised when they enter the sublimation distance (white dashed circle). The Roche radius is indicated with a red dotted circle, just outside the sublimation zone. \label{fig:emission_t_geometry}}
\end{figure*}
\begin{figure}
\centering
\includegraphics[width=\hsize]{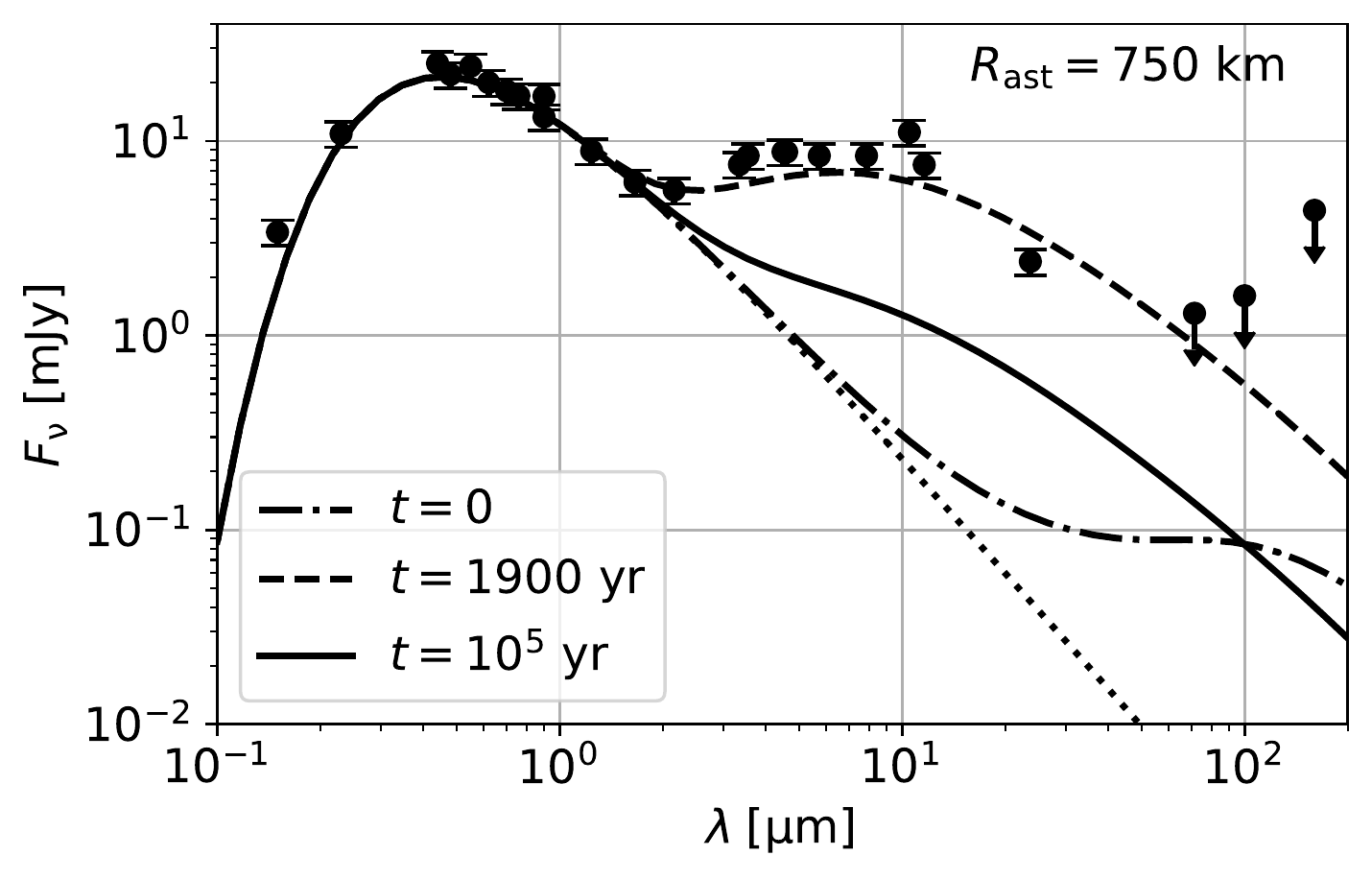} 
\caption{Simulated emission spectra from a 750 km asteroid disruption around G29-38 computed by our model (see text), followed by collision-less and optically thin orbital contraction via PR drag. The dotted line indicates the contribution from the star, with the dash-dotted ($t=0$), dashed ($t=1900 \; \mathrm{yr}$) and solid ($t=10^5 \; \mathrm{yr}$) lines including the emission from the fragments at different times. The points indicate the emission from the system as observed in several surveys (largely compiled by \citet{Farihi2014}, original references in the text). \label{fig:emission_t_spectrum}}
\end{figure}

\subsection{Infrared emission from a collision-less tidal disc}\label{sect:IR_excess_t}
Before we calculate the infrared emission by dust grains from collisional grind-down, we first evaluate how much emission is produced just after a tidal disc forms and how this changes as fragments circularise. The emission produced by newly formed tidal discs was suggested by \citet{Nixon2020} as the source of observed infrared excesses based on normalised emission profiles.

As explained in Sect. \ref{sect:stream_morphology}, the fragments are initially placed on a range of orbits depending on the size of the asteroid, that share their pericentre but differ in their apocentre (see Fig. \ref{fig:new_orbits}). In this calculation, we simplify the situation somewhat and assume that the fragments initially occupy a single orbital ring ($a_0, e_0$), given by the original orbit of the asteroid. Along this ring, the fragments experience different temperatures, which we approximate with radiative equilibrium:
\begin{equation}\label{eq:T_eq}
    T(r) = T_\mathrm{WD} \left(\frac{r}{R_\mathrm{WD}}\right)^{-\frac{1}{2}}.
\end{equation}
For simplicity, we ignore the super-heating of the smallest particles in this model \citep{Chiang1997, Rafikov2012}. The Planck emission per unit mass ($\mathcal{F}_\mathrm{\nu}$) depends on the size of the fragments, which determines their ratio of area ($A$) to mass ($M$):
\begin{subequations}
\begin{align}
    \mathcal{F}_\mathrm{\nu} (R,T) &= \frac{A B_\mathrm{\nu}(T)}{M} \\
    &= \frac{3B_\mathrm{\nu}(T)}{\rho_\mathrm{frag}R},
\end{align}
\end{subequations}
where $B_\mathrm{\nu}$ is the Planck function at a given frequency $\nu$. In this simplified collision-less test case, fragments evolve towards the star via PR drag as a function of time, with their semi-major axes and eccentricities slinking at rates given by Eqs. \ref{eq:da_dt} and \ref{eq:de_dt} \citep{Veras2015a, Veras2015b}. At any time $t$, fragments with different sizes $R$ have already shifted away from their initial orbit and are now located at their own orbits ($a(t),e(t)$), which are most contracted for the smallest fragments. We take a linear grid of angular bins with size $d\theta$ between 0 and $2\pi$, where fragments spend a time $dt = d\theta / \dot{\theta}(\theta)$. The average emission from a particular orbit per unit mass ($\mathcal{F}_\mathrm{\nu, orbit}$) is then weighed by the time spent at given angle $\theta$ as:
\begin{equation}
    \mathcal{F}_\mathrm{\nu, orbit}(R,t) =
    \frac{1}{P(R,t)} \sum_{\theta=0}^{2\pi} \mathcal{F}_\mathrm{\nu}(R,t,\theta) \frac{d\theta}{ \dot{\theta}(\theta)},
\end{equation}
where $P$ is the fragment's orbital period. The total emission ($F_\mathrm{\nu, PR}$) from the asteroid fragments is given by the sum over all the rings occupied by the different fragment sizes at time $t$:
\begin{equation}
    F_\mathrm{\nu, PR} (t) = \sum_{R_\mathrm{min}}^{R_\mathrm{max}} \mathcal{F}_\mathrm{\nu, orbit}(R,t) \frac{dM}{dR}(R) dR,
\end{equation}
where $dM/dR$ is defined by the fragment size distribution. As explained in Sect. \ref{sect:size_dist}, we take a truncated power law with the default exponent $\alpha=3.5$. The definite lower limit is set by the blow-out size (Eq. \ref{eq:R_blow}) and the upper limit is set at 1 km, corresponding to a rough indication of the breakup limit from the tidal force (Eq. \ref{eq:R_max}). 

We visualise the results by calculating the expected emission around G29-38, a 11240 K DAZ white dwarf at 17.5 pc with a well-documented infrared excess \citep{Xu2018}, first identified by \citet{Zuckerman1987}. We compare our results to multi-wavelength fluxes, for which we take a supplemented version of the sample compiled by \citet{Farihi2014} with observations from \citet{Tokunaga1990}, \citet{Reach2005}, and \citet{Farihi2008}. For our comparison here, we disrupt a large, 750 km asteroid with initial semi-major axis at 3 AU and let the fragments circularise in a collision-less, optically thin manner via PR drag. Without collisions, this scenario produces a peak accretion rate just above $10^9$ g/s that occurs around 1900 yr after the disruption event. We show the geometry of the emitting fragments around the white dwarf in Fig. \ref{fig:emission_t_geometry}, plotted at three different times at an emission wavelength of 10 $\mathrm{\mu m}$. Panel a ($t=0$) corresponds to the moment after disruption, assuming that the fragments are spread uniformly over the orbit, which is known to only require a few orbits \citep{Veras2014, Malamud2020a, Li2021}. Panel b is plotted at 1900 yr, just before the first fragments sublimate and accrete, which happens when their pericentre crosses the sublimation distance (white dashed circle), assumed to be located at 1500 K, corresponding to the temperature where the vapor pressure of silicates becomes significant and they begin to sublimate in vacuum \citep{Lieshout2014}. Panel (c) corresponds to $10^5$ yrs after the disruption event, when fragments smaller than 100 $\mathrm{\mu m}$ have already accreted. In all three cases, the non-circular nature of the material is evident. Most emission at 10 $\mathrm{\mu m}$ occurs near the orbit's pericentre, or just away from it in the fragment's most eccentric initial state. Because the fragments enter the star's sublimation radius before they are entirely circular, the central gap has the form of an ellipse rather than a circle. The eccentricity of this ellipse increases for higher stellar temperatures and for lower sublimation thresholds.

In Fig. \ref{fig:emission_t_spectrum}, we plot the emission spectrum of the fragments at the same snapshots in time. The first thing to note is that the emission varies greatly with time. The initial fragment orbits have semi-major axes similar to that of the original asteroid (3 AU here), and, therefore, have long orbital periods with most of their time spent at a distant apocentre. As a result, the emission from the orbit's pericentre is initially minimal, with emission only picking up at longer ($\sim 100 \; \mathrm{\mu m}$) wavelengths. Based on these calculations, we argue that newly formed tidal discs provide insufficient infrared emission to explain the observed excesses or variation, as was suggested by \citet{Nixon2020}. We find that the emission from the disc does rapidly increases as the fragments circularise, peaking in this collision-less test case at the moment that the smallest fragments begin to reach the star. The emission then gradually decreases again as the smallest fragments reach the white dwarf, with the emission spectrum maintaining a similar shape. During the circularisation process, the effective temperature of the emission slightly increases over time as fragments contract their orbits and spend more time at strongly irradiated locations.

\begin{figure}
\centering
\includegraphics[width=0.8\hsize]{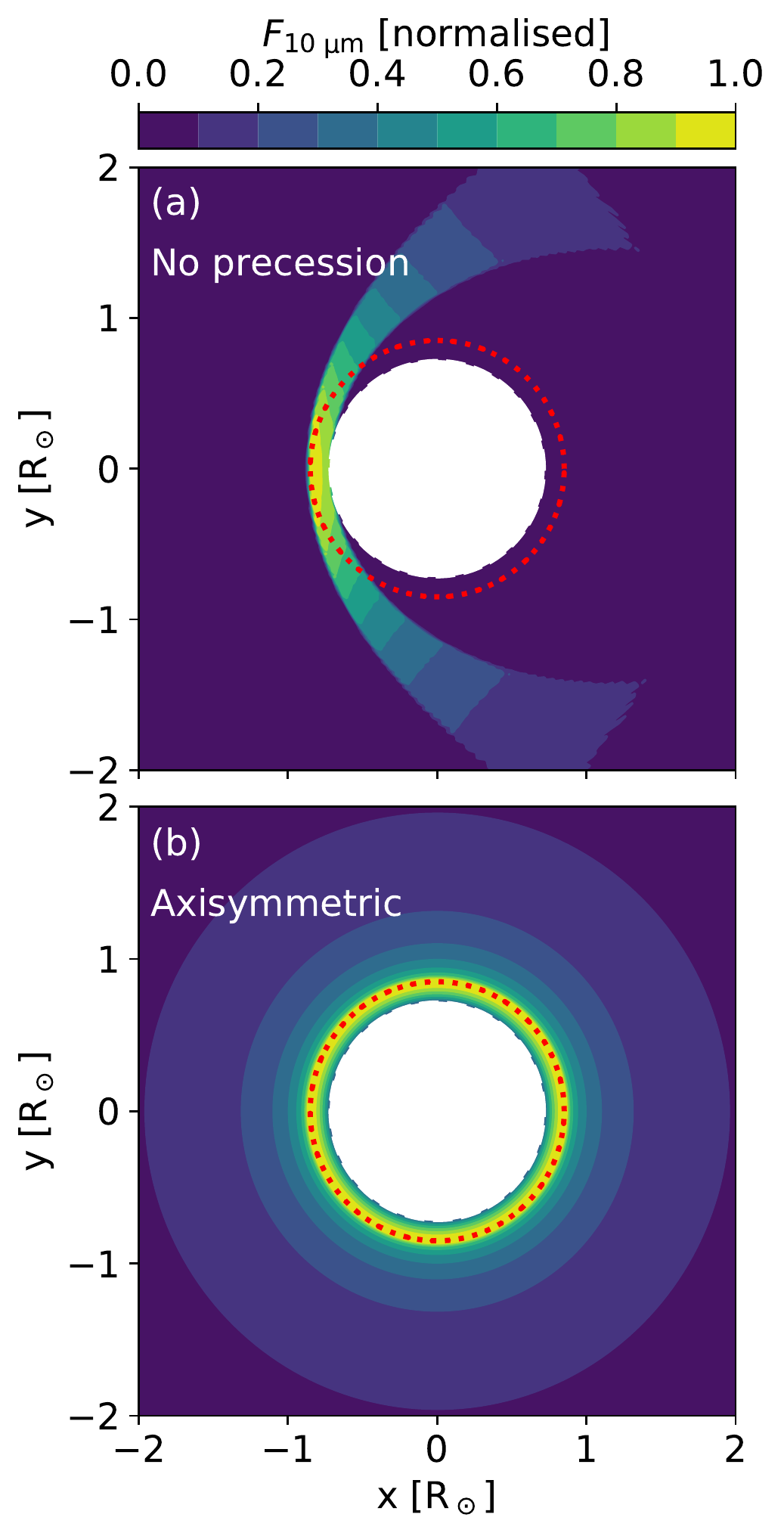} 
\caption{Simulated emission per unit surface area at 10 $\mathrm{\mu m}$ from constant dust production and accretion at a rate of $10^7$ g/s via PR drag, computed assuming optically thin properties. The top panel (a) shows the resulting emission without apsidal precession and the bottom panel (b) shows the axisymmetrically averaged emission. The disc contains an elliptical inner gap because fragments are not yet fully circularised when they enter the sublimation zone (white dashed circle). The Roche radius is indicated with a red dotted circle, just outside the sublimation zone. When axisymmetrically averaged, the structure becomes visible as a ring-like structure mostly contained within the Roche radius. \label{fig:emission_av_geometry}}
\end{figure}
\begin{figure}
\centering
\includegraphics[width=\hsize]{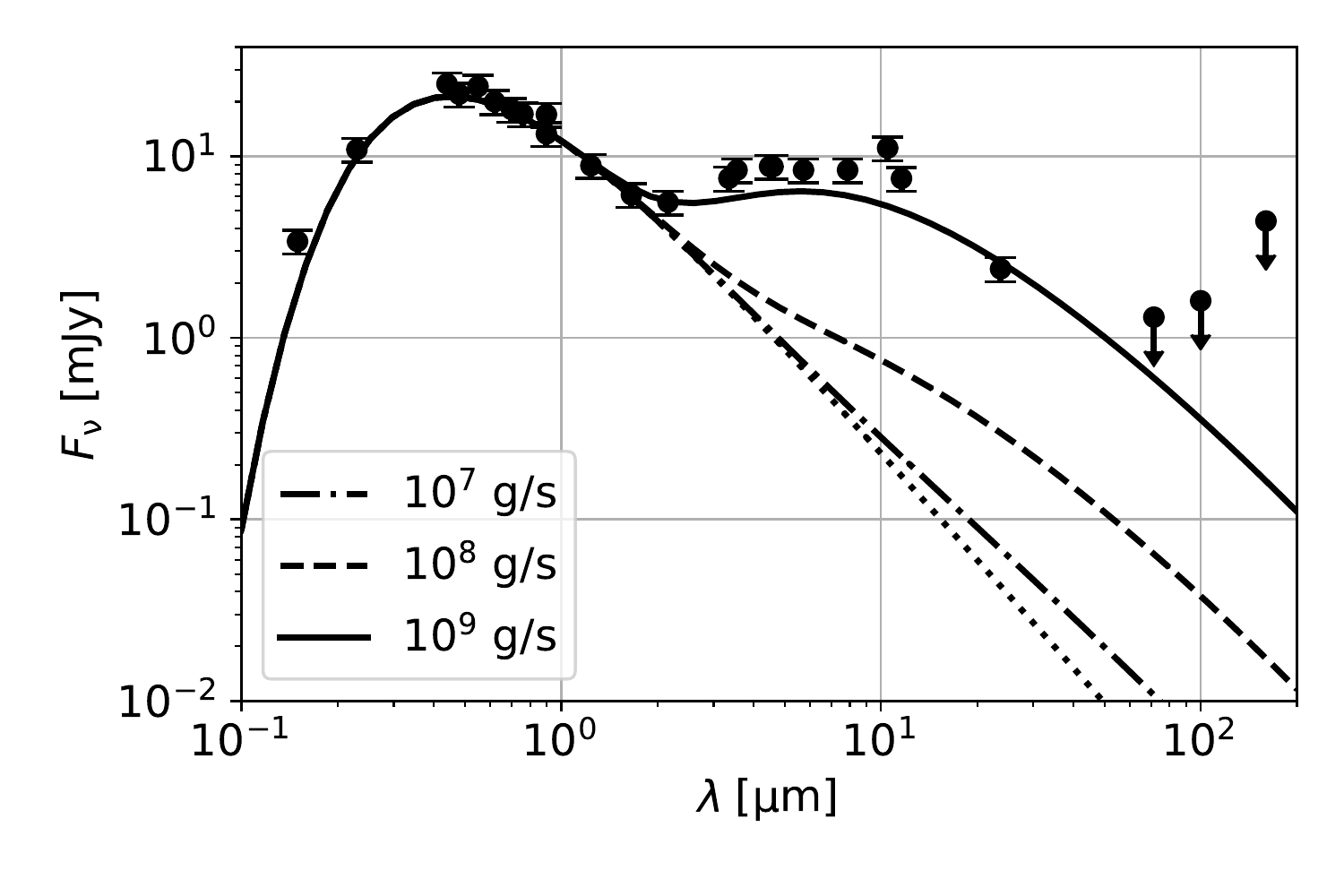} 
\caption{Simulated emission spectra of G29-38, with dust accretion via PR drag at three different rates. The dotted line indicates the contribution from the star, with the dash-dotted ($10^7$ g/s), dashed ($10^8$ g/s) and solid ($10^9$ g/s) lines including the emission from circularising dust at different accretion rates. The points indicate the observed emission from the system in several surveys (largely compiled by \citet{Farihi2014}, original references in the text). \label{fig:emission_av_spectra}}
\end{figure}

\subsection{Emission from collisional dust production}\label{sect:IR_excess_av}
We can slightly modify the calculation of the previous subsection to calculate the emission produced by the accretion of small dust that is produced via collisional grind-down of larger fragments. With the same assumption of optically thin contraction via PR drag, we can calculate the average emission per unit mass ($\bar{\mathcal{F}}_\mathrm{\nu}$) of a fragment from the moment of its production to its accretion by averaging over the different orbits ($a(t),e(t)$) that a dust grain occupies on its trajectory towards the star:
\begin{equation}
    \bar{\mathcal{F}}_\mathrm{\nu}(R) = \frac{1}{t_\mathrm{acc}(R)}\sum_{t'=0}^{t_\mathrm{acc}(R)} \mathcal{F}_\mathrm{\nu, orbit}(R,t') dt'(R,t').
\end{equation}
The time intervals $dt'$ during which the different orbital bins are occupied follow from Eqs. \ref{eq:da_dt}, \ref{eq:de_dt}.
We again sum the contributions of different fragments in the size distribution to obtain the total emission ($\bar{F}_\mathrm{\nu, PR}$), but now for a given accretion rate rather than at a particular time:
\begin{equation}
    \displaystyle\bar{F}_\mathrm{\nu, PR} (\dot{M}) = \sum_{R_\mathrm{min}}^{R_\mathrm{max}} \bar{\mathcal{F}}_\mathrm{\nu}(R) t_\mathrm{acc}(R) \frac{d\dot{M}}{dR}(R) dR,
\end{equation}
which, for a given accretion rate, notably yields a result that is independent of the fragment size distribution if the fragments are modeled as black bodies in thermal equilibrium \footnote{The PR accretion time scales as $t_\mathrm{acc}\propto R$, while the area-to-mass ratio scales as $A/M\propto R^{-1}$. These contributions cancel out, meaning that the emission in a steady-state of dust production and accretion by PR drag is independent of R if the particles are modeled as perfect absorbers/emitters.}. We show the geometry of the emission in Fig. \ref{fig:emission_av_geometry}, which, as expected, shows a similarly asymmetric shape as the time evolution discussed in the previous subsection, with most of its emission originating from a narrow band near the pericentre of the orbit. If the collisions are induced by the differential apsidal precession of fragment orbits, the actual geometric shape of the disc will be different. We approximate this state in panel (b) of the same figure, where we neglect potential collisional circularisation from inelastic collisions, yielding an axisymmetrically averaged structure when the fragment orbits are completely differentially precessed. The pericentre now appears as a bright narrow band up to the Roche radius, surrounded by a less luminous zone.

We plot the emission spectrum, which is not affected by asymmetric averaging, in Fig. \ref{fig:emission_av_spectra}, corresponding to accretion onto G29-38 at three different rates. Because the emission is proportional to the assumed accretion rate, the implication of this model is that an excess in the system's infrared emission is visible whenever the PR accretion rate is sufficiently high. While such an excess has indeed been observed for some rapidly accreting systems, the majority of polluted white dwarfs do not follow this trend. Hence, we must conclude that even if PR drag can supply high accretion rates of small dust, the majority of systems likely accrete their material in a different manner. In the next subsection, we will evaluate the optical depth of the accretion via PR drag, which provides an additional constraint to the dust accretion process via discs with small inclinations.

\subsection{Radial optical depth of the tidal debris disc}\label{sect:optical_depth}
In the previous calculations regarding accretion via PR drag, we assumed that the dust and fragments around the white dwarf form an optically thin disc. Here, we investigate the validity of this assumption and compute the optical depth in the radial direction, where the disc is most opaque. We follow a similar procedure as in the previous subsection and begin by computing the radial optical depth contribution per unit mass ($\mathcal{T}$) as a function of the true anomaly for a given orbit:
\begin{subequations}
\begin{align}
    \mathcal{T}(R,t,\theta) &= \frac{\pi R^2}{M H dl} \frac{dt}{P} \\
    &= \frac{3}{4\rho_\mathrm{frag}RH(R,t,\theta)P(R,t)v(R,t,\theta)},
\end{align}
\end{subequations}
where $P(R,t)$ is the period of a fragment at its given orbit. For this calculation, we assume the same constant inclination imparted at the moment of breakup of $i = R_\mathrm{ast}/r_\mathrm{B}$, which is around $10^{-4}$ for a 100 km asteroid (see Sect. \ref{sect:stream_morphology}). To obtain the optical depth contribution per unit mass ($\overline{\mathcal{T}}(R,\theta)$) for a fragment size bin, we again sum over the different orbital rings that are occupied during the orbital contraction, evaluated separately for every angle:
\begin{equation}
    \overline{\mathcal{T}}(R,\theta) = \frac{1}{t_\mathrm{acc}(R)}\sum_{t'=0}^{t_\mathrm{acc}(R)} \mathcal{T}(R,t',\theta) dt'(R,t').
\end{equation}
The total optical depth is then determined by the sum over the size distribution:
\begin{equation}
    \displaystyle\tau_\mathrm{PR}(\dot{M},\theta) = \sum_{R_\mathrm{min}}^{R_\mathrm{max}} \overline{\mathcal{T}}(R,\theta) t_\mathrm{acc}(R) \frac{d\dot{M}}{dR}(R) dR.
\end{equation}
We are primarily interested in estimating the optical depth at pericentre because that is where the fragments are exposed to the most stellar light, even when corrected for travel time \citep{Veras2015a,Veras2015b}. When we perform the calculation with the parameters of G29-38 for the star and assume a 100 km asteroid, we find that the disc becomes radially opaque at its pericentre when the accretion rate is increased beyond $10^7$ g/s. This result is not altered much when the material in the disc is axisymmetrically averaged, meaning that the optical depth at pericentre is comparable to the average across the disc. 

While stellar rays from more vertical directions can still reach the dust when it is radially opaque, our calculation suggests that the assumption of perfect optically thin accretion begins to fail at higher accretion rates. If the inclination of the dust remains small, the amount of stellar light that can reach the dust grains declines and circularisation rates slow down (see also the models by \citealt{Rafikov2011a, Rafikov2011b}). However, if the inclination of the tidal disc does increase from its initial value, for instance due to interactions with neighboring planets \citep{Li2021}, the disc can remain entirely optically thin and high dust accretion rates by PR drag are possible. This scenario could explain the few cases where infrared excesses are observed. Finally, several white dwarfs show the remarkable combination of both rapid ongoing accretion and no infrared excess. In the scenario of grind-down and dust accretion, we find that this requires faster dust circularisation and accretion than is possible by PR drag alone, as was also suggested by \citet{Bonsor2010}. This finding points to the importance of additional circularision processes that could involve gas drag \citep{Malamud2021} or the recently suggested mechanism of Alfv\'en-wave drag \citep{Zhang2021}.

\section{Discussion}\label{sect:discussion}
In this work, we studied how material accretes onto white dwarfs from their surrounding planetary systems and how this relates to observational quantities. Our baseline scenario begins with the tidal disruption of an asteroid close to the white dwarf, which forms a highly eccentric tidal debris disc. The fragment orbits are then perturbed via various processes, including differential apsidal precession, causing the larger fragments to collide on their eccentric orbits until only dust remains. In the final stage, the dust accretes onto the star by drag forces. Our suggested scenario can produce accretion rates as high as those observed ($\gtrsim 10^{11}$ g/s) from the disruption of $\gtrsim 200$ km asteroids. Both the presence and the absence of infrared emission can be explained depending on the rate of dust in-spiral and accretion, with drag rates faster than PR-drag, such as via additional gas or Alfv\'en-wave drag required to explain the absence of infrared emission in the majority of white dwarfs. 

\begin{figure}
\centering
\includegraphics[width=0.992\hsize]{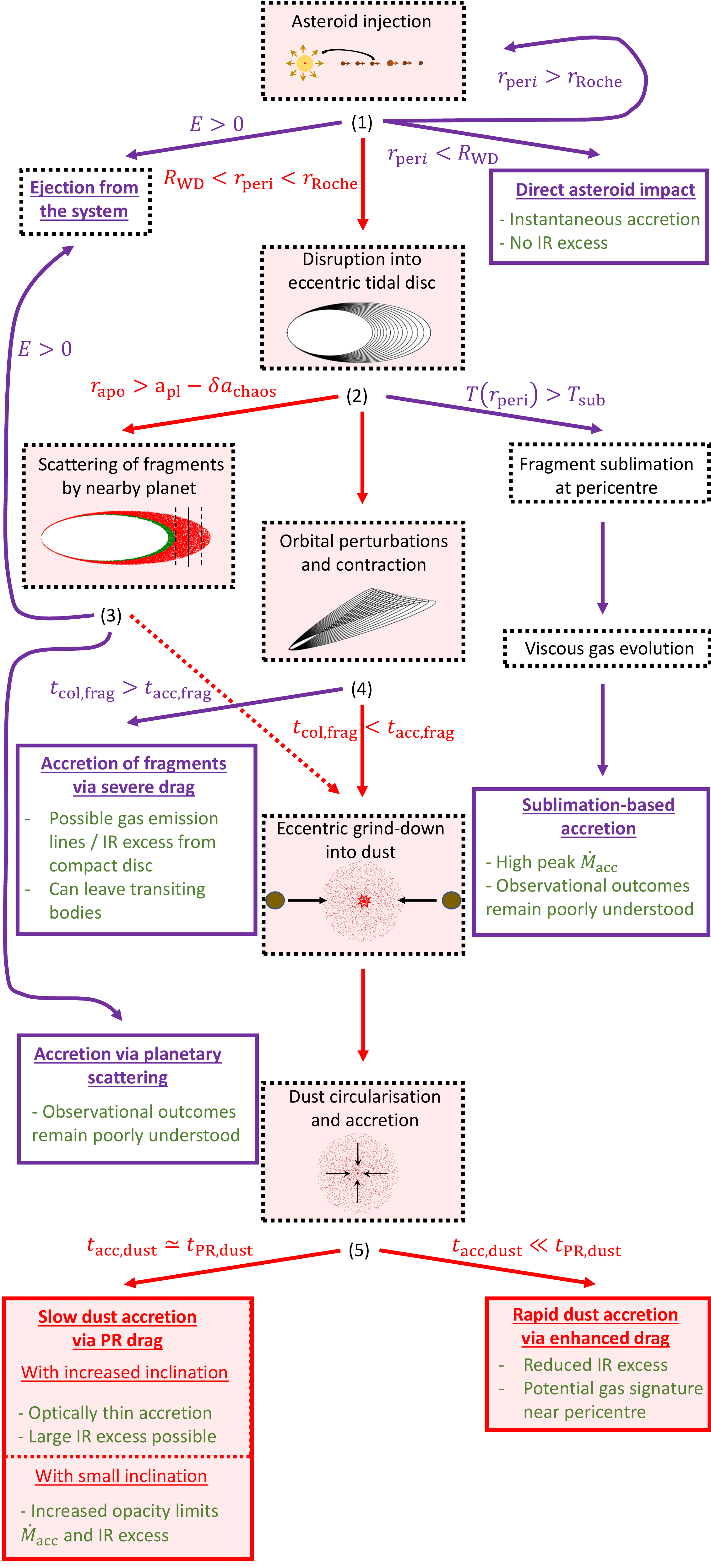}
\caption{Road-map outlining potential routes for planetary material to arrive in the atmospheres of white dwarfs. Our suggested main route (red arrows and boxes) begins with the injection of an asteroid into the stellar Roche radius, followed by a tidal disruption event, orbital perturbations, collisional grind-down, and finally dust accretion. Alternative accretion channels are shown in purple with physical selection criteria at five numbered points. The detectable characteristics of these different accretion channels are listed in green, provided that they are sufficiently well-constrained. \label{fig:overview}}
\end{figure}

Nevertheless, we propose that no single accretion scenario explains the pollution of all white dwarfs. A range of different accretion channels are likely applicable depending on the properties of both the central star and the accreted object. In this discussion, we will present a selection of different routes to white dwarf pollution, which are summarised in the form of a road-map in Fig. \ref{fig:overview}. We visualise this road-map as a series of forks that split into different accretion channels according to specified physical criteria. We also discuss the observational characteristics that belong to each of these channels wherever they are sufficiently well understood, noting that further detailed modelling is required in many cases. After presenting our road-map, we will discuss the limitations of our calculations relating to our suggested main path and suggest areas for improvement in our model.

\subsection{Fork (1): Direct asteroid impact, ejection or tidal debris disc formation}
The first step towards white dwarf pollution starts with mass loss from the central star, which widens the chaotic zone around planets \citep{Bonsor2011, Mustill2018} and can destabilise tightly-packed planetary systems \citep{Debes2002, Maldonado2020, Maldonado2021}. Nearby asteroids become subject to scattering from close encounters or strong perturbations in mean motion resonances. In principle, scattering events can have four possible outcomes, which we show at the first fork (1) of Fig. \ref{fig:overview}. Most commonly, the asteroid's eccentricity or semi-major axis are only altered slightly and the asteroid continues on its way until it is scattered again. In a chain of scattering events, the asteroid can eventually attain such a high eccentricity that it enters the Roche radius of the star and it disrupts into an eccentric tidal disc (red arrow). This corresponds to our suggested baseline model. Alternatively, the asteroid could either be scattered outwards and enter the influence of outer planets, become completely unbound from the system, or directly hit the surface of the white dwarf if its pericenter distance becomes sufficiently small. 

This last possibility of a direct asteroid impact is worth mentioning as a separate channel of accretion, studied in detail by \citet{Brown2017} and \citet{Mcdonald2021}. It is the simplest method of mass accretion, as material almost instantly enters the star's photosphere. This near-instant accretion prevents any detectable infrared excess and also restricts the detection window of the pollution itself to a few sinking timescales of metals in the photosphere, making direct asteroid impacts unlikely to be detected in young DAZ stars. In any case, direct impacts should be rare events. Not only is the Roche radius significantly larger than the white dwarf itself, but \citet{Veras2021a} show that most low mass (terrestrial) planets provide small eccentricity kicks that marginally push asteroids into the white dwarf's tidal disruption zone. 

\subsection{Fork (2): Sublimation, continued scattering by a planet or orbital perturbations and collisions}\label{sect:discussion_scenarios}
We continue along the main channel of our road-map (visualised as red in Fig. \ref{fig:overview}) with the formation of an eccentric tidal disc. Following the asteroid's disruption, its fragments can evolve in a number of ways as indicated at the second fork (2). If the temperature at the disc's pericenter exceeds the sublimation threshold, its fragments quickly turn into gas. This is always the expected outcome around hot white dwarfs ($\gtrsim 20.000$ K), which are sufficiently luminous that rocky material begins sublimating at distances outside the star's Roche radius \citep{Bonsor2017}. However, in rarer cases, fragments can also sublimate around less luminous stars if their pericenter lies deep in the Roche sphere or when the fragments are composed of volatile components, which always sublimate within the Roche radius \citep{Steckloff2021}. Although it is clear that fragment sublimation leads to a distinct channel of accretion, further work is required to determine the full details. If the subsequent viscous evolution of the gas is sufficiently rapid, the gas can quickly accrete onto the star after it is produced, leading to a scenario of pure gas accretion with potentially detectable gas emission lines but no infrared emission. If the gas does not circularise sufficiently within a single orbit, it likely re-condenses on its way back towards apocentre, in which case dust exterior to the Roche limit could produce detectable infrared emission. In any case, this scenario is likely characterised by a high peak accretion rate as small fragments quickly sublimate and the presence of gas only adds to the circularisation and accretion speeds of remaining solids.

For those tidal discs where sublimation does not occur, even at pericentre, we consider two further possibilities. If the asteroid was scattered by a planet, this planet could potentially re-scatter the disrupted material. This would occur for those fragments whose apocentre continues to approach that of the planet (See sect. \ref{sect:planet_scattering}). Otherwise, the fragments will evolve according to further orbital perturbations, including apsidal precession, leading to collisions. We separate these two scenarios because they potentially lead to different observational outcomes, as discussed next.

\subsection{Fork (3): Outcome of continued scattering by a planet}
We visualise the possible outcomes of continued scattering after fork (3) of Fig. \ref{fig:overview}. Planets can either scatter fragments outwards, such that they are ejected, inwards, such that they graze the star with reduced pericenter, or just perturb their orbits, leading to further collisional evolution. The likely outcome depends in part on the size of the asteroid and on its semi-major axis. Since larger asteroids and those that originate from an outer belt disrupt into wider tidal discs, strong scattering by a planet is unlikely for most of their fragments, such that we envisage continued collisional grind-down as the most likely pathway in these cases. This preference towards collisional evolution is further amplified for large asteroids due their fragments shorter collisional time-scales. For smaller asteroids, collisions require more time and many fragments remain on planet-crossing orbits, making them instead susceptible to continued strong scattering. We predict that as the sublimation zone is significantly larger than the white dwarf, and the bodies are deep in the white dwarf's potential, inward scattering of fragments typically leads to their sublimation rather than a direct impact. In this sense, the accretion channel via planetary scattering might proceed similarly to the sublimation-based channel mentioned earlier. Understanding the full details will require further work, where a detailed understanding of gaseous evolution and condensation on the highly eccentric orbits will be crucial.

\subsection{Fork (4): Rapid circularisation or collisional grind-down}
We continue our suggested main road at fork (4) of Fig. \ref{fig:overview} with the evolution of fragments that are not sublimated form the heat of the central star nor scattered by a planet. These fragments nevertheless have their orbits perturbed via various processes (see Sect. \ref{sect:perturbations}) until they either lose enough angular momentum to accrete onto the star intact, or until they collide with a different fragment. Whether their orbits can fully contract before a catastrophic collision occurs, depends mainly on the size of the asteroid progenitor and the time required for circularisation. Larger asteroids produce more fragments and lead to faster collisions. Speed is key here and PR drag, the most suggested process for angular momentum loss, is clearly too slow. It was already shown by \citet{Veras2015b} that PR drag takes too long to accrete fragments above $\sim 10$ cm before the star cools down, let alone before they collide with other fragments. In our analysis of the tidal and material forces involved, we estimate that the upper limits to fragment sizes lie much higher, around 100 m - 10 km depending mainly on the strength of the asteroid (see Sect. \ref{sect:disruption}). 

There are, however, other processes that can contract orbits at a much greater pace than PR drag. One of these is the drag induced by fragment interactions with regions around the star that contain high concentrations of either gas or dust grains, for instance in the form of a compact disc that formed from in prior accretion event. In a recent work, \citet{Malamud2021} showed that drag at the fragment's pericentre can even circularise km-sized bodies in several orbits, provided that a large second object already formed a massive pre-existing disc around the star. If this disc only exists for a limited duration of time and is thick enough to circularise all but the very largest fragments, remaining km-sized boulders could survive into a new environment where they are relatively safe from collisional grind-down. Although its statistical significance has yet to be evaluated, this could form one channel to generate transiting material, as is observed around some systems \citep{Vanderburg2015, Manser2019b, Vanderbosch2020, Vanderbosch2021}. The main argument against the significance of this accretion channel is the lack of observations of gaseous emission lines or large IR excesses belonging to the required pre-existing disc.

\subsection{Fork (5): A link between the dust circularisation speed and infrared excess}
In our suggested main channel, we continue with the collisional grind-down of fragments into dust. In this final phase, the speed of the dust circularisation leads to a split in possible observational outcomes. With our optically thin emission model, we show that slow dust circularisation in a sufficiently inclined disc via PR drag leads to detectable infrared excesses at higher accretion rates ($\gtrsim 10^7$ g/s). This scenario could explain the minority of systems that show significant infrared excesses. If the inclination of the dust instead remains equal to the tiny value imparted during the tidal disruption event, the work done by stellar light becomes limited by the radial optical depth of the grains and the circularisation of the shaded dust grains slows down, ultimately limiting accretion onto the star and limiting the IR excess.

However, the scenario of slow dust circularisation via PR drag cannot be used to explain the majority of systems that show no detectable infrared excess, even at high accretion rates. We suggest, therefore, that PR drag is not the only force that drives dust circularisation around most polluted white dwarfs. As was earlier suggested by \citet{Bonsor2010}, other drag forces - likely involving gas - are likely to play a key role. If the time required to circularise dust grains is reduced sufficiently by the gas drag, their accretion can occur without the accumulation of high grain abundances around the central star. In this manner, different circularisation speeds of dust around different stars could break the proportionality between accretion rate and infrared excess. It is possible that the small quantities of gas required to accelerate the accretion of dust grains are readily produced in the grind-down process itself. Indeed, Doppler tomography shows that some systems contain gas near the Roche radius, likely as a consequence of collisional production \citep{Manser2016a,Steele2021}.

\subsection{Main road: model caveats and improvements}
Having discussed conceivable alternative paths to white dwarf pollution along with their physical selection criteria, we finally evaluate the model limitations of our suggested main road to accretion. The main uncertainty in the first stage relates to the fragment size distribution. As discussed in Sect. \ref{sect:disruption}, the upper limit of the distribution is limited by our uncertain knowledge of the material strength. At the lower end, it is not clear whether the smallest grains are indeed produced in disruption events, and the slope is very poorly constrained. These factors severely limit the quantitative conclusions of our grind-down calculations, which are strongly related to our assumed size distribution. 

Despite the great uncertainties relating to the fragment sizes, we argue that it is worthwhile to evaluate the grind-down to examine the process at order-of-magnitude scale and to investigate the trends and biases it involves. Our simple grind-down model of Sect. \ref{sect:collisional_model} should be interpreted in this way, rather than as an attempt to predict exact accretion rates. Firstly, it is based only on angular differences induced by differential apsidal precession and does not include any other perturbing processes. Clearly these other perturbing forces would play an important role. However, the precession rates for differential apsidal precession are analytically known, such that these could be readily incorporated. We hypothesise that the general trends in collisional rates will follow a similar form. Secondly, the calculation only tracks catastrophic collisions and does not track the full collisional evolution of child orbits. This can be justified tentatively by the faster collisions of smaller fragments that result from the collisions but it remains an important limitation of the model. A more self-consistent evolution is possible to simulate in theory but is numerically difficult considering the extreme eccentricity ($\sim 0.999$) of the fragment orbits. Given our limited knowledge of the fragment size distribution, we did not consider that such a model would significantly improve our understanding of the processes involved. Although it could still be worthwhile to develop such a detailed model in the future, its predictive power will remain limited as long as the fragment size distribution after a tidal breakup event remains poorly constrained.

Similarly, our calculation of the infrared excess in the dust accretion stage is done in a simplified manner with the main goal to elucidate trends and show the two-dimensional morphology of the system rather than to predict precise excesses. Most importantly, we only performed calculations in the limiting case that the accreting dust is optically thin, whereas the discs become radially optically thick if the rate of dust production is greater than $\sim 10^7$ g/s and the discs inclination remains small. Although it is clear that the inclination is directly linked to the observational outcome of dust accretion, further work is required to study how it evolves after the tidal disc is formed.

\section{Summary and conclusions}\label{sect:conclusions}
The main aim of this paper is to produce a road-map illustrating several potential routes for white dwarf pollution and to link these paths to observational outcomes (see Fig. \ref{fig:overview}). Our main route begins with the tidal disruption of a scattered asteroid and the formation of a highly eccentric tidal disc, followed by the collisional grind-down of fragments, which finally circularise and accrete onto the star as dust due to drag forces. Alternatives to this standard pathway for white dwarf pollution include a) the direct accretion of scattered asteroids/fragments of asteroids onto the white dwarf, b) the sublimation of dusty material and the accretion of gas, or c) the rapid circularisation of fragments via pre-existing compact discs. While accretion likely proceeds through a combination of these channels, the alternative a) is statistically very unlikely to occur, even with a planet re-scattering fragments of a disrupted asteroid. Channel b) will occur only for material scattered sufficiently close to the hottest white dwarfs, while c) occurs only following previous disruption events.

Here we present detailed calculations of our suggested main road to white dwarf pollution. Our work includes simulations of collisional grind-down due to differential precession as well as a model for the infrared excess of the dust that is produced. Our main findings are that:
\begin{enumerate}
    \item The size distribution of fragments in a tidal disruption event around a white dwarf can range as many as 10 orders of magnitude. The smallest bound fragments are no smaller than the limit set by radiation pressure at 0.1-10 $\mathrm{\mu m}$ (Fig. \ref{fig:blowout}), while fragments as large as 100 m-10 km also survive the tidal disruption depending on their material strength (Fig. \ref{fig:Strengths}). In the absence of a pre-existing compact disc or intense radiation, these larger fragments must be ground down before they can circularise and accrete by drag forces.
    
    \item Large asteroids produce more fragments when they disrupt, causing rapid collisional grind-down and generating short and intense bursts of dust production, whereas smaller asteroids grind down over longer periods of time. If subsequent dust accretion is fast, this biases observations to detect on-going accretion at intermediate rates by smaller asteroids (Fig. \ref{fig:cumulative_P}). Rare peaks in accretion rates from large asteroids are short-lasting and only probable to be detected in the atmospheres of DBZ stars with longer sinking time-scales (Fig. \ref{fig:cumulative_single}).
    
    \item Optically thin dust discs produce large amounts of infrared emission when their accretion rate exceeds $10^7$ g/s. However, in order to remain completely optically thin at these high accretion rates, the inclination of the dust grains must be substantially increased beyond the tiny value imparted at the moment of tidal disruption. Infrared excesses at high accretion rates can be avoided by more rapid dust circularisation, for instance via enhanced drag due to the presence of gas near the disc’s pericentre.
\end{enumerate}
%

\section*{Data availability}
The simulation data that support the findings of this study are available upon request from the corresponding author, Marc G. Brouwers.

\section*{Acknowledgements}
We would like to thank Laura Rogers, Andrew Buchan and Dimitri Veras for useful discussions that helped shape this paper. In particular, we are grateful to Elliot Lynch for useful input regarding the discussion of coherent apsidal precession and for suggesting the possibility of direct fragment scattering by a planet. We are also very grateful for the initial work on this project by Tom Callingham during his Part III Project at the University of Cambridge. We thank Nicholas Ballering for kindly providing us an updated multi-wavelength sample of G29-38. Marc G. Brouwers acknowledges the support of a Royal Society Studentship, RG 160509. Amy Bonsor is grateful to the Royal Society for a Dorothy Hodgkin Fellowship.

\bibliographystyle{mnras}
\bibliography{circularisation}
%

\end{document}